\documentclass[a4paper,11pt]{article}

\usepackage{graphicx}
\usepackage{subfig}
\usepackage{color}
\usepackage{amsfonts,amssymb,latexsym,amsmath}
\usepackage[section]{placeins}
\usepackage{a4wide}
\usepackage{bm}
\usepackage{array} 
\usepackage{hyperref}
\usepackage{epstopdf}
\usepackage{cite}
\usepackage{tikz}
\usepackage[compat=1.1.0]{tikz-feynman}
\usepackage{simplewick}
\usepackage{feynmp-auto}
\usepackage{slashed}
\usepackage{bbold}
\usepackage{multirow}
\usepackage{cleveref}
\begin{document}

\title{\bf Semileptonic $B_c$ decays to P-wave charmonia in the light-cone QCDSR}


\author{
Tarik Akan$^{a, b}$\thanks{tarik.akan@bozok.edu.tr}~~, Elif Cincioglu$^{c}$\thanks{elifcincioglu@gmail.com}~~, Altug Ozpineci$^{c}$\thanks{ozpineci@metu.edu.tr}~~ and Adnan Tegmen$^{b}$\thanks{tegmen@science.ankara.edu.tr}\\
{\small \it $^a$Physics Department, Yozgat Bozok University,  66100 Yozgat, Turkey}\\
{\small \it $^b$Physics Department, Ankara University,  06100 Ankara, Turkey}\\
{\small \it $^c$Physics Department, Middle East Technical University,  06800 Ankara, Turkey}
}

\maketitle

\begin{abstract}
$B_c$ mesons are laboratories for probing heavy quark physics. Furthermore, they decay into charmonia and hence their decays can be used to analyze possible exotic charmonium. In this work, the semileptonic decays of $B_c$ mesons into P wave charmonia are analysed using light cone QCD sum rules (LCSR). The distributions amplitudes for 
the charmonia are taken from a quark model computation. The obtained decay rates for the ground state and excited charmonia are compared with the results found in the
literature.
\end{abstract}


\newpage

\section{Introduction}

In the past several decades, many particle physicists devoted themselves to explain composite quark systems, the so-called hadrons, and checked predictions of quantum chromodynamics. Many models are developed in perturbative and non-perturbative domains for this purpose. During the last few decades, heavy flavoured hadron research has become very popular in hadron physics since many charmonium-like states have been discovered at B-factories. Among these states, some of them may not be simple quark and anti-quark states. They could be more complex structures, the so-called exotic states, such as tetraquarks, pentaquarks, meson molecules, etc. Therefore, there is not a certain hadron structure. The possible meson states which include the b quarks, are the hidden b quark states, the so-called $\Upsilon$ ($b\bar{b}$) and the open b quark states, which are called B mesons ($b\bar{q}$, where $q=u,d,s,c$).  
Among the B mesons, $B_c$ meson ($b\bar{c}$) is a very interesting particle due to its two different kind of heavy quark constituents and final states of some of its decays contain a $c\bar{c}$ pair \cite{LHCb:2014rck, LHCb:2017vlu}. The decays and spectroscopy of $B_c$ mesons are described by various QCD models, which can be tested at B-factories. Due to small velocities of constituent quarks in heavy quarkonia and $B_c$ mesons, the mesons can be treated as non-relativistic systems.  Many non-relativistic potential models claim that the b and $\bar{c}$ are tightly bound in a very compact system and have a rich spectroscopy of excited states \cite{Godfrey:1985xj}. 
Investigation of these excited states also increases the popularity of $B_c$ states. Since the first successful observation of $B_c$ in CDF at Tevatron, Fermilab in the interaction $B_c\rightarrow J/\psi\, l\, \bar{\nu_l}$ \cite{CDF:1998axz}, there has been remarkable progress in the study of semileptonic and nonleptonic B meson decays  \cite{Aliev:1998mq, Aliev:1998ka, Aliev:2006vs, Aliev:1999tg, Ghahramany:2008tz, Azizi:2008vv, Azizi:2008vy, Azizi:2007jx, Azizi:2008tw,Kang:2018jzg}. Besides the OZI suppressed decay channels, in which the $b$ and $\bar{c}$ annihilate, $B_c$ meson can also decay due to the decay of one of its constituents. If the final quark is a light quark, the final state is a D or a B meson depending on which constituent decays.
If the b quark decays into a c quark, than the final state contains a charmonium. 
Its weak decay channel shows us a large branching ratio to final states containing a $J/\psi$. 
It has been pointed out that \cite{Aliev:1998mq, Aliev:1998ka, Azizi:2008tw, Azizi:2008vy, Du:1988ws, Gershtein:1994jw, Qiao:2012hp}, due to its unique nature, 
the decays of $B_c$ meson can be used to extract the magnitudes and phases of CKM matrix elements.
Additionally, if the final state contains a $c\bar c$ pair, the decay can be used to probe hidden charm exotic states.
For example, the produced $c\bar{c}$ state can couple directly or through a $D\bar{D}^*/\bar{D}D^*$ loop to X(3872)\cite{Esposito:2014rxa, Canham:2009zq, Sun:2015uva,Kang:2016jxw}. 

Due to the non-perturbative nature of QCD \cite{Shifman:1978bx, Alkofer:2000wg, Marciano:1977su, Bigi:1992su}, to investigate the semileptonic decays of $B_c$ meson, a non-perturbative method is needed. 
QCD sum rules (QCDSR) and its variations like the Light Cone QCDSR  have been used successfully to analyze the properties of hadrons. In the sum rules analysis,
the properties of low lying states can be easily extracted but the question of studying the properties of excited states is still an open problem.
For this reason, e.g. in \cite{Azizi:2009ny, Kiselev:1999sc,Ebert:2002pp}, only the decays of $B_c$ into ground state P-wave charmonia has been analyzed. 
In X(3872), if it contains any charmonium components, it is expected to be the first radial excitation \cite{Cincioglu:2016fkm}. Hence
if $B_c$ meson decays will be used to probe the structure of possible exotic states like X(3872) it is important to study decays into final states that contain
the radial excitations. 
In \cite{Brodsky:1997de}, light cone DAs are related to the quark model wave functions. In \cite{Olpak:2016wkf,Hwang:2009cu} this relation is used to calculate the DAs describing the ground state
and radially excited P-wave charmonia. In this work, in order to study the decays of $B_c$ mesons into P-wave charmonia, the results of \cite{Olpak:2016wkf,Hwang:2009cu} are used.

In \cref{sec:SR} the procedure is explained in more detail.  Important parameters of LCSR and a mass prediction for $B_c$ are discussed and presented in the \cref{sec:analysis}.
Then the obtained numerical results for decay rate and branching fractions, including their excited states are presented in \cref{sec:decay} and compared with existing results in the literature.
Finally, our conclusion is given in the \cref{sec:conc}.

\section{Form Factor Calculation with LCSR}
\label{sec:SR}

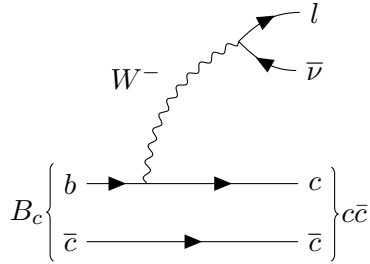
\begin{figure}
\centering
\begin{tikzpicture} 
\begin{feynman}
\vertex(a4) {\(b\)};
\vertex[right=1cm of a4] (a5); 
\vertex[right=2cm of a5] (a6) {\(c\)};

\vertex[below=2em of a4] (b5) {\(\overline c\)};
\vertex[below=2em of a6] (b6) {\(\overline c\)};

\vertex[above=of a6] (c1) {\(\overline \nu\)}; 
\vertex[above=2em of c1] (c3) {\(l\)}; 
\vertex at ($(c1)!0.5!(c3) - (1cm, 0)$) (c2);

\diagram* { (a4)--[fermion](a5)--[fermion](a6), (b5)--[fermion](b6),
(c1) -- [fermion, out=180, in=-45] (c2) -- [fermion, out=45, in=180] (c3),
(a5) -- [boson, bend left, edge label=\(W^{-}\)] (c2), };

\draw [decoration={brace}, decorate] (b5.south west) -- (a4.north west) node [pos=0.5, left] {\(B_{c}\)};
\draw [decoration={brace}, decorate] (a6.north east) -- (b6.south east) node [pos=0.5, right] {\(c \bar c\)};

\end{feynman} 
\end{tikzpicture}

\caption{Feynman diagram corresponding to the semileptonic decay $B_c\rightarrow (c\bar{c})\,l\,\bar{\nu}$}
\label{fig:fig1}
\end{figure}

 
The $B_c\rightarrow (c\bar{c})\,l\,\bar{\nu_l}$ decays proceed via the $b\rightarrow c$ transition at the quark level \cref{fig:fig1}. 
This transition is described by the effective Hamiltonian: \cite{Azizi:2009ny,Ebert:2010zu,Issadykov:2017wlb}
\begin{equation}
	H_{eff}=\dfrac{G_F}{\sqrt{2}}V_{cb}\,J_{lep}^\mu J_\mu^{had},
\label{eq:eq1}
\end{equation}
where $G_F$ is the Fermi constant, $V_{cb}$ is the CKM matrix element for the $b \rightarrow c$ transition, and the currents are given by $J_{lep}^\mu=\bar{l}\gamma^\mu(\mathbb{1}-\gamma_5)\nu_l$ and $J_\mu^{had}=\bar{c}\gamma_\mu(\mathbb{1}-\gamma_5)b$.
The transition matrix element of the decay can be obtained by sandwiching the effective Hamiltonian between corresponding initial and final states, which becomes after factoring out the hadronic and leptonic parts
\begin{equation}
	M=\dfrac{G_F}{\sqrt{2}}V_{cb}\,\langle l\bar{\nu}_l|\bar{l}\,\gamma_{\mu}\,(\mathbb{1}-\gamma_5)\,\nu_l|0\rangle \langle c\bar{c}(p^\prime)|\bar{c}\,\gamma_{\mu}\,(\mathbb{1}-\gamma_5)\,b|B_{c}(p)\rangle,
	\label{eq:eq3}
\end{equation}
where $B_c$ is the initial pseudoscalar $B_c$ meson and the final $c \bar c$ state can be a scalar (S), axial-vector(A) or a tensor(T) charmonium.
The matrix elements $\langle c\bar{c}\vert \bar{c}\,\gamma_{\mu}\,(\mathbb{1}-\gamma_5)\,b\vert B_c(p)\rangle$ for each possibility can be parameterized in terms of form factors as follows \cite{Azizi:2009ny, Ebert:2010zu,Leljak:2019eyw}:  
\begin{align}
	\langle S^n(p^\prime)| \bar{c}\,\gamma_{\mu}\,(\mathbb{1}-\gamma_5)\,b|B_c(p)\rangle &=f_1^n(q^2)(p^{\prime}+p)_{\mu}+f_2^n(q^2)(p-p^{\prime})_{\mu},
	\label{eq:eq4}
\\
%
\label{eq:eq5}
	\langle A^n(p^\prime,\varepsilon)| \bar{c}\,\gamma_{\mu}\,(\mathbb{1}-\gamma_5)\,b|B_c(p)\rangle &=\dfrac{i f_V^n(q^2)}{(m_{B_c}+m_{A})}\varepsilon_{\mu\rho\alpha\beta}\varepsilon^{*\rho}p^\alpha p^{\prime \beta}\\ \nonumber
	&+i\bigg[ f^n_{V_0}(q^2)(m_{B_c}+m_{A})\varepsilon_\mu^{*}-\dfrac{f^n_{V_+}(q^2)}{(m_{B_c}+m_{A})}(\varepsilon^*p)(p^{\prime}+p)_\mu\\ \nonumber
	&-\dfrac{f^n_{V_-}(q^2)}{(m_{B_c}+m_{A})}(\varepsilon^*p)(p-p^{\prime})_\mu\bigg],
\\
\label{eq:eq6}
    \langle T^n(p^\prime,\varepsilon)| \bar{c}\,\gamma_{\mu}\,(\mathbb{1}-\gamma_5)\,b|B_c(p)\rangle &=\dfrac{2 i f_T^n(q^2)}{m_{B_c}+m_{T}}\varepsilon^{\mu\rho\alpha\beta}\varepsilon_{\rho\sigma}^*\dfrac{p^{\sigma}}{m_{B_c}}p_{\alpha}p_{\beta}^{\prime}\\\nonumber
	&+i\bigg[f^n_{T_0}(q^2)(m_{B_c}+m_{T})\varepsilon^{*\mu\alpha}\dfrac{p^{\alpha}}{m_{B_c}} \\\nonumber
	&-f^n_{T_+}(q^2)p^{\mu}\varepsilon_{\alpha\beta}^*\dfrac{p^{\alpha}p^{\beta}}{m_{B_c}^2}-f^n_{T_-}(q^2)p^{\prime\mu}\varepsilon_{\alpha\beta}^*\dfrac{p^{\alpha}p^{\beta}}{m_{B_c}^2}\bigg],
\end{align}
where $n$ is the radial excitation quantum number of the charmonium, $f_1^n$ and $f_2^n$ are the form factors for the scalar; $f^n_V$,
$f^n_{V_0}$, $f^n_{V_+}$ and $f^n_{V_-}$ are the form factors, $m_A$ is the mass and $\epsilon_\mu$ is the polarization vector of the axial-vector;  
$f^n_T$, $f^n_{T_0}$, $f^n_{T_+}$, and $f^n_{T_-}$ are the form factors, $m_T$ is the mass and $\epsilon_{\mu\nu}$ the polarization tensor of the tensor P-wave charmonium.  

To obtain the form factors within the LCSR framework, a suitably chosen correlation function is studied. In this work, the correlator is chosen to be:
\begin{equation}
	\Pi_{\mu}=i\int d^4 x \,e^{i p x}\langle c \bar c(p')| T\{ j^W_{\mu}(0) j_{B}^{\dagger}(x) \} |0\rangle.
	\label{eq:eq7}
\end{equation}
where $j_{\mu}^W=\bar c \gamma_\mu (\mathbb{1} - \gamma_5) b$ is the weak current responsible for the $b \rightarrow c$ transition and $j_{B}^{\dagger}$ is an operator that can create a $B_c$ meson from the vacuum.
In this work $j_B$ is chosen as $j_B = \bar c \gamma_5 b$. Hence the correlation function can be written as
\begin{equation}
	\Pi_{\mu}=i\int d^4 x \,e^{i p x}\langle c\bar{c}(p^\prime)|T\{\bar{c}(0)\,\gamma_{\mu}(\mathbb{1}-\gamma_5)\,b(0)\bar{b}(x)\,\gamma_5\,c(x)\}|0\rangle.
	\label{eq:eq8}
\end{equation}

The main idea behind LCSR method is to write the correlation function \cref{eq:eq8} in terms of both the QCD degrees of freedom and light cone DAs, so called the QCD representation, and also in terms of the properties of the hadrons, so called the phenomenological representation.
In the kinematical region $p^2 > 0$, the correlation can be written in terms of hadronic parameters. To obtain this hadronic representation, a resolution of identity in terms of the hadronic states is inserted between the interpolating currents. In the hadronic representation, the correlation 
function can be written as:
\begin{equation}
	\Pi_{\mu}^{phen.}(p^2)=\sum_h \dfrac{\langle c\bar{c}(p^{\prime})|\bar{c}\,\gamma_{\mu}\,(\mathbb{1}-\gamma_5)\,b|h\rangle\langle h|\bar{b}\,\gamma_5\,c|0\rangle}{m_h^2-p^2} +\cdots ,
	\label{eq:eq9}
\end{equation}
where the sum is over all single hadron states, h, for which $\langle h \vert \bar b \gamma_5 c \vert 0 \rangle \neq 0$, and $\cdots$ represents the contributions from multi hadronic states and continuum. Note that in eq. (\ref{eq:eq9}), there is no sum over $c \bar c$ states as these states are parametrized by their distribution amplitudes. This allows one to study the properties of excited $c \bar c$ states if their distribution amplitudes are known.
Separating out the contribution of the lowest mass state, the correlation function can be written as
\begin{equation}
	\Pi_{\mu}^{phen.}(p^2)=\dfrac{\langle c\bar{c}(p^{\prime})|\bar{c}(0)\,\gamma_{\mu}\,(\mathbb{1}-\gamma_5)\,b(0)|B_c(0^-)\rangle\langle B_c(0^-)|\bar{b}(0)\,\gamma_5\,c(0)|0\rangle}{m_{B_c}^2-p^2}.
	\label{eq:eq10}
\end{equation}
where the contribution from the higher $B_c$ mesons and the continuum are omitted. 
The matrix element $\langle B_c(0^-)\vert \bar{b}(0)\,\gamma_5\,c(0)\vert 0\rangle$ can be written in terms of the leptonic decay constant of $B_c$ as: 
\begin{equation}
	\langle B_c(0^-)|\bar{b}(0)\,\gamma_5\,c(0)|0\rangle=-\dfrac{i f_{B_c}m_{B_c}^2}{m_b+m_c},
	\label{eq:eq11}
\end{equation}
where $f_{B_c} $ is the leptonic decay constant  of $B_c$ meson.

Using the definition of the remaining matrix elements in terms of the form factors given in \cref{eq:eq4,eq:eq5,eq:eq6} , the phenomenological representation of the correlation function becomes:

- For spin-0 charmonium states;
\begin{align}
	\Pi_{\mu}^{phen}(p^2)&=\dfrac{i}{m_{B_c}^2-p^2}(f^n_1(q^2)(p^{\prime}+p)_{\mu}+f^n_2(q^2)(p-p^{\prime})_{\mu})\dfrac{f_{B_c}m_{B_c}^2}{m_b+m_c}
	\nonumber \\ 
&	\equiv i (p'+p)_\mu \Pi^{phen}_{f^n_1} +i (p-p')_\mu \Pi^{phen}_{f^n_2}
		\label{eq:eq12}	
\end{align}

- For spin-1 charmonium states;
\begin{align}
	\Pi_{\mu}^{phen}(p^2)&=\dfrac{i}{m_{B_c}^2-p^2}\bigg\{\dfrac{i f^n_V(q^2)}{(m_{B_c}+m_{A})}\varepsilon_{\mu\rho\alpha\beta}\varepsilon^{*\rho}p^\alpha p^{\prime \beta}\\ \nonumber
	&+i\bigg[ f^n_{V_0}(q^2)(m_{B_c}+m_{A})\varepsilon_\mu^{*}-\dfrac{f^n_{V_+}(q^2)}{(m_{B_c}+m_{A})}(\varepsilon^*p)(p^{\prime}+p)_\mu\\ \nonumber
	&-\dfrac{f^n_{V_-}(q^2)}{(m_{B_c}+m_{A})}(\varepsilon^*p)(p-p^{\prime})_\mu\bigg]\bigg\}\dfrac{f_{B_c}m_{B_c}^2}{m_b+m_c}\\ \nonumber
	&\equiv \varepsilon_{\mu\rho\alpha\beta}\varepsilon^{*\rho}p^\alpha p^{\prime \beta}\Pi_{f^n_V}^{phen}+\varepsilon_{\mu}^{*} \Pi_{f^n_{V_0}}^{phen}-(\varepsilon^*p)(p^{\prime}+p)_\mu\Pi_{f^n_{V_+}}^{phen}\\ \nonumber
	&-(\varepsilon^* p)(p-p^{\prime})_\mu\Pi_{f^n_{V_-}}^{phen}
\label{eq:eq13}	
\end{align}

- For spin-2 charmonium states;
\begin{align}
	\Pi_{\mu}^{phen}(p^2)&=\dfrac{i}{m_{B_c}^2-p^2}\bigg\{ \dfrac{2 i f^n_T(q^2)}{m_{B_c}+m_{T}}\varepsilon^{\mu\rho\alpha\beta}\varepsilon_{\rho\sigma}^*\dfrac{p^{\sigma}}{m_{B_c}}p_{\alpha}p_{\beta}^{\prime}\\\nonumber
	&+i\bigg[f^n_{T_0}(q^2)(m_{B_c}+m_{T})\varepsilon_{\mu\alpha}^*\dfrac{p^{\alpha}}{m_{B_c}}\\\nonumber
	&-f^n_{T_+}(q^2)(p+p^\prime)_{\mu}\varepsilon_{\alpha\beta}^*\dfrac{p^{\alpha}p^{\beta}}{m_{B_c}^2}-f^n_{T_-}(q^2)(p-p^\prime)_{\mu}\varepsilon_{\alpha\beta}^*\dfrac{p^{\alpha}p^{\beta}}{m_{B_c}^2}\bigg]\bigg\}\dfrac{f_{B_c}m_{B_c}^2}{m_b+m_c}\\ \nonumber
	&\equiv \varepsilon^{\mu\rho\alpha\beta}\varepsilon_{\rho\sigma}^* p^\sigma p_{\alpha}p_{\beta}^{\prime}\Pi_{f^n_T}^{phen}+ \varepsilon_{\mu\alpha}^* p^\alpha\Pi_{f^n_{T_0}}^{phen}-(p+p^\prime)_{\mu}\varepsilon_{\alpha\beta}^* p^\alpha p^\beta\Pi_{f^n_{T_+}}^{phen}\\ \nonumber
	&-(p-p^\prime)_{\mu}\varepsilon_{\alpha\beta}^*p^\alpha p^\beta \Pi_{f^n_{T_-}}^{phen}
	\label{eq:eq14}	
\end{align}
where  the coefficients of certain Dirac structures in \cref{eq:eq4,eq:eq5,eq:eq6} are defined as:
\begin{equation}
    \Pi^{phen}_{f^n_X} = \frac{i f^n_X}{m_{B_c}^2-p^2} \kappa_{f^n_X}
\end{equation}
where $f^n_X$ are formfactors and $\kappa_{f^n_X}$ are constants for each form factor $f^n_X$ are defined as
\begin{align}
    \kappa_{f^n_1} = \kappa_{f^n_2} &=\dfrac{f_{B_c}m_{B_c}^2}{m_b+m_c}\\
    \kappa_{f^n_V} &= -\dfrac{1}{m_{B_c}+m_A} \dfrac{f_{B_c}m_{B_c}^2}{m_b+m_c}\\
    \kappa_{f^n_{V_0}} &= -(m_{B_c}+m_A) \dfrac{f_{B_c}m_{B_c}^2}{m_b+m_c}\\
    \kappa_{f^n_{V_+}} = \kappa_{f^n_{V_-}} &= -\dfrac{1}{m_{B_c}+m_A} \dfrac{f_{B_c}m_{B_c}^2}{m_b+m_c}\\
    \kappa_{f^n_T} &= -\dfrac{2}{(m_{B_c}+m_T)} \dfrac{f_{B_c}m_{B_c}}{m_b+m_c}\\
    \kappa_{f^n_{T_0}} &=- (m_{B_c}+m_T)\dfrac{f_{B_c}m_{B_c}}{m_b+m_c}\\
    \kappa_{f^n_{T_+}} =\kappa_{f^n_{T_-}} &= - \dfrac{f_{B_c}}{m_b+m_c}
\end{align}
The various form factors $f^n_X$ can be obtained in terms of the QCD parameters once $\Pi_{f^n_X}$ is expressed in terms of these parameters and the two representations of $\Pi_{f^n_X}$ are matched.

In the kinematical region when $p^2$ is both large and negative, i.e. the deep Euclidean region, the correlation function can be calculated in terms of QCD parameters. To calculate $\Pi_\mu$ in terms of the QCD parameters, the first step is to contract the b quark fields and \cref{eq:eq8} is expressed as:

\begin{align}
	\Pi_{\mu}(p^2)&=i\int d^4 x \,e^{i p x}\langle c\bar{c}(p^\prime)|\bar{c}(0)\,\gamma_{\mu}(\mathbb{1}-\gamma_5)\,\contraction{}{b}{(0)}{\bar b}b(0)\bar{b}(x),\gamma_5\,c(x)|0\rangle\\\nonumber
	&=i\int d^4 x \,e^{i p x}\langle c\bar{c}(p^\prime)|\bar{c}(0)\,\gamma_{\mu}(\mathbb{1}-\gamma_5)\,iS_b(x)\,\gamma_5\,c(x)|0\rangle,
	\label{eq:eq15}
\end{align}
where the heavy quark propagator is defined as
\begin{equation}
	S_Q(x)=\dfrac{m_Q^2}{4\pi^2}\bigg[\dfrac{i\slashed{x}}{(-x^2)}K_2(m_Q\sqrt{-x^2})+\dfrac{1}
{(\sqrt{-x^2})}K_1(m_Q\sqrt{-x^2})\bigg].
\label{eq:eq16}
\end{equation}
where the $K_n(m_Q \sqrt{-x^2})$ ($n=1$ or $2$) are the modified Bessel functions. Using the Fierz identity, this expression of the correlation function can be written as:

\begin{align}
	\Pi_{\mu}(p^2)&=\sum_\Gamma \dfrac{i}{4}\int d^4\, x e^{i p x}\langle c\bar{c}(p^{\prime})|\bar{c}(0)\,\Gamma\, c(x)|0\rangle\, Tr[\,\gamma_{\mu}(\mathbb{1}-\gamma_5)S_b(x)\gamma_5\,\Gamma\,]
	\label{eq:eq17}
\end{align} 
where $\Gamma$ is summed over the Dirac basis for gamma matrices: $\{$$\mathbb{1}$, $\gamma_5$, $\gamma_{\mu}$, $i \gamma_{\mu}\gamma_5$, $\frac{\sigma_{\mu\nu}}{\sqrt2}$$\}$. 
The matrix elements appearing in \cref{eq:eq17} can be written in terms of the distribution amplitudes of the corresponding $c\bar c$ state under study.
The leading twist contributions have been calculated in \cite{Olpak:2016wkf, Hwang:2009cu}. Note that, due to kinematical enhancements, higher twist DA's can receive
large contribution that can be expressed in terms of the leading twist DAs using Wandruza-Wilczek like relations. 
The non-zero matrix elements can be written as \cite{Hwang:2009cu}:
\begin{equation}
	\langle S^n(P)|\bar{c}(0)\gamma^\mu c(x)|0\rangle=f^n_S\int_0^1 du\,e^{-i\bar{u}\,p\,x}\bigg[p^\mu\phi^n_{S}(u)+x^\mu\dfrac{m_S^2}{2px}g_s(u)\bigg],
	\label{eq:eq18}
\end{equation}
\begin{align}
	\langle A^n(P,\varepsilon_{\lambda=0})|\bar{c}(0)\gamma^\mu\gamma_5c(x)|0\rangle =-if^n_A\,m_A\int_0^1 du\,e^{-i\bar{u}\,p\,x}\bigg\{&p^\mu \dfrac{\varepsilon x}{p x}\phi^n_{A_\parallel} (u)
	+\varepsilon_{\perp}^\mu g_{A_\perp}(u)\\\nonumber&-x^\mu\dfrac{\varepsilon x}{2(px)^2}m_A^2 g_{A_3}(u)\bigg\},
	\label{eq:eq19}
\end{align}
\begin{align}
	\langle A^n(P,\varepsilon_{\lambda=\pm1})|\bar{c}(0)\sigma^{\mu\nu}\gamma_5c(x)|0\rangle=f^n_{A_\perp}\int_0^1 du\,e^{-i\bar{u}\,p\,x}\bigg\{&(\varepsilon_\perp^\mu p^\nu-\varepsilon_\perp^\nu p^\mu)\phi^n_{A_\perp} (u)\\\nonumber&+(p^\mu x^\nu-p^\nu x^\mu)\dfrac{m_A^2 \varepsilon x}{(px)^2}h_{A_\parallel}(u)\\\nonumber&+(\varepsilon_\perp^\mu x^\nu-\varepsilon^\nu x^\mu)\dfrac{m_A^2}{2px}h_{A_3}(u)\bigg\},
	\label{eq:eq20}
\end{align}
\begin{align}
	\langle T^n(P,\varepsilon_{\lambda=0})|\bar{c}(0)\gamma^\mu c(x)|0\rangle=f^n_T\,m_T^2\int_0^1 du\,e^{-i\bar{u}\,p\,x} \bigg\{&p^\mu \dfrac{\varepsilon^{\bullet\bullet}}{(p x)^2}\phi^n_{T_\parallel} (u)+\dfrac{\varepsilon_\perp^{\mu\bullet}}{p x} g_{T_\perp}(u)\\\nonumber&-x^\mu\dfrac{\varepsilon^{\bullet\bullet}}{2(px)^3}m_T^2 g_{T_3}(u)\bigg\},
	\label{eq:eq21}
\end{align}
\begin{align}
	\langle T^n(P,\varepsilon_{\lambda=\pm1})|\bar{c}(0)\sigma^{\mu\nu}c(x)|0\rangle=-if^n_T\,m_T\int_0^1 du\,e^{-i\bar{u}\,p\,x}\bigg\{&\dfrac{\varepsilon_{\perp}^{\mu\bullet}p^\nu-\varepsilon_{\perp}^{\nu\bullet}p^\mu}{p x}\phi^n_{T_\perp} (u)\\\nonumber&+(p^\mu x^\nu-p^\nu x^\mu)\dfrac{m_T^2\varepsilon^{\bullet\bullet}}{(px)^3}h_{T_\parallel}(u)\\\nonumber&+(\varepsilon^{\mu\bullet}_\perp x^\nu-\varepsilon^{\bullet\nu}_\perp x^\mu)\dfrac{m_T^2}{2(px)^2}h_{T_3}(u)\bigg\},
	\label{eq:eq22}
\end{align}
where
\begin{equation}
    p^\mu=P^\mu-x^\mu\dfrac{m_H^2}{2Px}
\end{equation}
where $m_H$ is the charmonium mass in question. The subleading twist DAs that receive large contributions are also included, $\bar{u}=1-u$ and S, A, T correspond to scalar, axial-vector and tensor, respectively. The large contributions to the subleading twist DAs can be calculated as \cite{Ball:1996tb,Braun:2000cs,Cheng:2010hn}:
\begin{equation}
    g_{S}(u)\simeq \dfrac{1}{2}\bigg[\int_0^u\, dv\dfrac{\phi_{S}(v)}{\bar{v}}+\int_u^1\,dv\dfrac{\phi_{S}(v)}{v}\bigg], \label{eq:ww1}
\end{equation}
\begin{equation}
    g_{A_\perp}(u)\simeq \dfrac{1}{2}\bigg[\int_0^u\, dv\dfrac{\phi_{A_\parallel}(v)}{\bar{v}}+\int_u^1\,dv\dfrac{\phi_{A_\parallel}(v)}{v}\bigg],
    \label{eq:ww2}
\end{equation}
\begin{equation}
    g_{T_\perp}(u)\simeq\int_0^u\, dv\dfrac{\phi_{T_\parallel}(v)}{\bar{v}}+\int_u^1\,dv\dfrac{\phi_{T_\parallel}(v)}{v},
    \label{eq:ww3}
\end{equation}
\begin{equation}
    h_{A_\parallel}(u)\simeq 2\bigg[\bar{u}\int_0^u\, dv\dfrac{\phi_{A_\perp}(v)}{\bar{v}}+u\int_u^1\,dv\dfrac{\phi_{A_\perp}(v)}{v}\bigg],
    \label{eq:ww4}
\end{equation}
\begin{equation}
    h_{T_\parallel}(u)\simeq (2u-1)\bigg[\int_0^u\, dv\dfrac{\phi_{T_\perp}(v)}{\bar{v}}+\int_u^1\,dv\dfrac{\phi_{T_\perp}(v)}{v}\bigg]. \label{eq:ww5}
\end{equation}
Note that in \cref{eq:ww1,eq:ww2,eq:ww3,eq:ww4,eq:ww5}, only contributions to DAs appearing in our final results are presented. The polarization vector of the axial-vector meson is separated into longitudinal and transverse parts as

\begin{equation}
	\varepsilon_\perp^\mu=\varepsilon^\mu-\varepsilon_\parallel^\mu,\qquad \varepsilon_\parallel^\mu=\dfrac{\varepsilon x}{p x}\bigg(p^\mu-x^\mu\dfrac{M_H^2}{2 p x}\bigg).
	\label{eq:eq23}
\end{equation}
On the other hand, for the tensor meson, polarization tensor is similarly described in the following way

\begin{equation}
	\varepsilon_\perp^{\mu\bullet}=\varepsilon^{\mu\bullet}-\varepsilon_\parallel^{\mu\bullet},\qquad \varepsilon_\parallel^{\mu\bullet}=\dfrac{\varepsilon^{\bullet\bullet}}{p x}\bigg(p^\mu-x^\mu\dfrac{M_H^2}{2 p x}\bigg),
	\label{eq:eq24}
\end{equation}
where $\varepsilon_{\mu\bullet}=\varepsilon^{\mu\nu}x_\nu$. It should be noted that there is no contribution to the matrix elements of mesons with polarizations $\lambda=\pm 2$ from the leading twist \cite{Braguta:2008qe, Wang:2009mi}. 
Depending on  the $C$ parities of the $c \bar c$ states, the distribution amplitudes are odd or even under the $u \leftrightarrow \bar u$ exchange. $\phi_S$, $\phi_{T_\parallel}$,and $\phi_{T_\perp}$ are odd under the $u\leftrightarrow\bar{u}$ transform since the scalar and tensor P-wave charmonia have positive C-parities.  However, there are two axial P-wave charmonia: one C-odd and the other C-even. 
In the case of C-even axial vector mesons, $\phi_{A_\parallel}$, $\phi_{A\perp}$ wave functions are even and odd, respectively and vice versa for the C-odd axial-vector mesons. 

Once these non-local matrix elements are defined, 
the correlation function can be expressed in terms of the distributions amplitudes after a straightforward computation. From this expression, 
taking the coefficients of various Dirac structures, one can obtain 
expression of the function $\Pi^{QCD}_{f_X}$ in terms of the QCD parameters.

The expressions for the function $\Pi_{f_X}$ obtained in the two kinematical regions, $p^2 > 0$ and $p^2$ in the deep Euclidean region, are matched using the spectral representation of the correlation function. In general, the correlation function, or more precisely the coefficient of any of the Dirac structures appearing in the correlation function
can be written as
\begin{align}
    \Pi(q^2,p^2) = \int_0^\infty ds \frac{\rho(q^2,s)}{s-p^2} + \mbox{polynomials in $p^2$.}
\end{align}
where $\rho$ is called the spectral density. To get rid of the polynomials 
and suppress the contribution of higher states and continuum, Borel transformation is carried out, after which one obtains:
\begin{align}
    f^n_X \kappa_{f^n_X} e^{-\frac{m_{B_c}^2}{M^2}} + \cdots = \int_0^\infty ds \rho^{QCD}_{f^n_X}(q^2,s) e^{-\frac{s}{M^2}}. 
\end{align}
where $\cdots$ represents the contribution of the higher states and continuum.
To model and subtract the contributions of higher states and the continuum, quark hadron duality is used. It is assumed that 
above a threshold value of $s_0$, $\rho^{phen}$ and $\rho^{QCD}$ are equal to each other:
\begin{equation}\label{eq:eq33}
    \rho_{f^n_X}^{\text{higher states}}(q^2,s)=\rho_{f^n_X}^{QCD}(q^2,s)\theta(s-s_0)
\end{equation}
where $s_0$ is called the threshold value.

Eventually, the light cone sum rules prediction for the form factors can be obtained from:
\begin{align}
    f^n_X(q^2)e^{-\frac{m_{B_c}^2}{M^2}} = \frac{1}{\kappa_{f^n_X}} \int_0^{s_0} ds \rho^{QCD}(q^2,s) e^{-\frac{s}{M^2}}
    \label{eq:eqsr}
\end{align}

The analytical results for the form factors are presented in \cref{app:fff}. 

\section{Numerical Analysis for Form Factors}
\label{sec:analysis}
In order to obtain the numerical values for the right hand side of \cref{eq:eqsr}, the values of the quark masses, explicit expression of LCDA's, 
the masses of the $c \bar c$ states and the mass of the $B_c$ meson are needed. 
For the quark masses, the following pole masses are used: \cite{Zyla:2020zbs,Narison:2019tym} $m_b=4.78$ GeV, $m_c=1.67$ GeV.
\begin{table}
\begin{center}
\begin{tabular}{|c|c|c|c|}
\hline 
Masses $(GeV)$           & $n=1$ & $n=2$ & $n=3$ \tabularnewline
\hline 
$m_{S}\, (^{3}P_{0})$ & $3.37$ & $3.88$ & $4.30$ \tabularnewline
\hline 
$m_{A}\, (^{3}P_{1})$ & $3.54$ & $3.97$ & $4.33$ \tabularnewline
\hline 
$m_{A}\, (^{1}P_{1})$       & $3.53$ & $3.96$ & $4.37$ \tabularnewline
\hline 
$m_{T}\, (^{3}P_{2})$ & $3.54$ & $3.98$ & $4.34$ \tabularnewline
\hline 
\end{tabular}
\caption{Quark model masses calculated for the first three levels of charmonia \cite{Olpak:2016wkf}}
\label{table:masses}
\end{center}
\end{table}
The distribution amplitudes for the ground state and the next two excited states of the P-wave charmonium are modeled in \cite{Olpak:2016wkf,Braguta:2008qe} as
\begin{align}
\phi_{odd}^{n=1}(u)&=a(1-u^2)^2\bigg[E_3\big[\frac{\beta}{1-u^2}\big]+b\, exp\big(-\frac{u^2}{c}\big)\bigg]  \label{eq:eq25} \\
\phi_{odd}^{n=2,3}(u)&=a\bigg[\frac{1}{1+\frac{(u^2-u_0^2)^2}{\sigma^2}}+b\, exp\big(-\frac{u^2}{c}\big)\bigg]exp\big(-\frac{\beta}{1-u^2}\big)	
\label{eq:eq26}
\end{align}

\begin{align}
\phi_{even}^{n=1}(u)&=a\,u(1-u^2)\bigg[exp\big(\frac{\beta}{1-u^2}\big)+b\, exp\big(-\frac{u^2}{c}\big)\bigg]  \label{eq:eq27}\\
\phi_{even}^{n=2,3}(u)&=-\frac{d}{du}\bigg\{a\bigg[\frac{1}{1+\frac{(u^2-u_0^2)^2}{\sigma^2}}+b\, exp\big(-\frac{u^2}{c}\big)\bigg]exp\big(-\frac{\beta}{1-u^2}\big)	\bigg\}
\label{eq:eq28}
\end{align}
where $a$, $b$, $c$, $\beta$, $\sigma^2$, and $u_0$ are numerical parameters whose values are presented in \cite{Olpak:2016wkf}. 
In general LCDA's are scale dependent. In the present work, the typical scale of the problem is a few $m_c$, but in \cite{Olpak:2016wkf}, the numerical parameters are given for $\mu=m_c$ and $\mu=\infty$. In the present work, the values of the numerical parameters at the scale $\mu=m_c$ are used. Changing the scale to $\mu=\infty$ modified the obtained results by less than 3\%. 

\begin{figure}	
	\subfloat[]{
  		\includegraphics[width=0.5\textwidth]{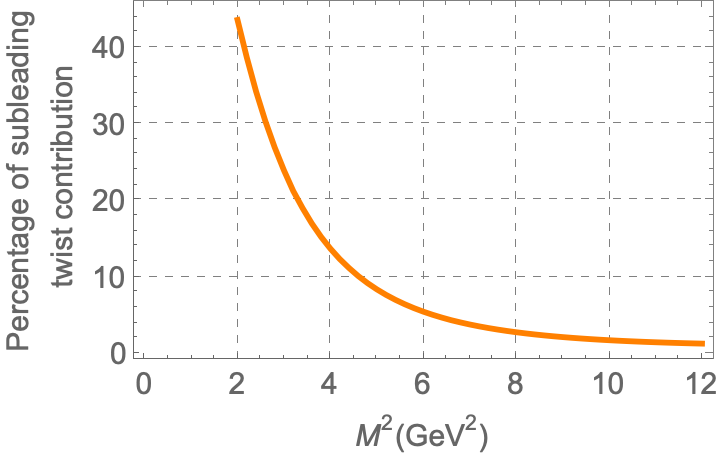}}
	\subfloat[]{
  		\includegraphics[width=0.5\textwidth]{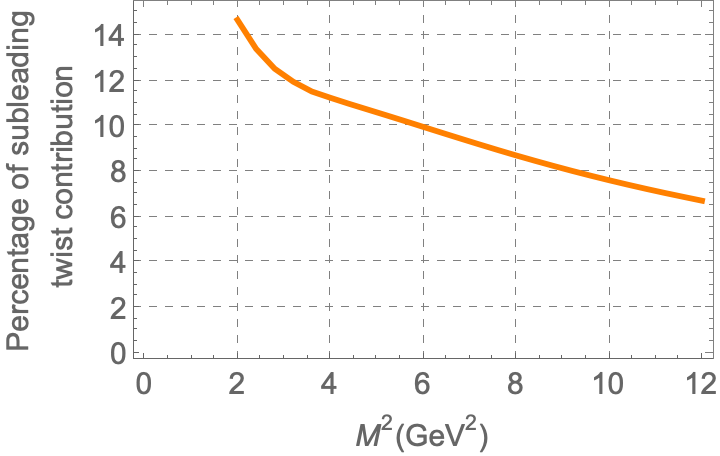}}\\
  	\caption{The percentage of subleading twist contribution-$M^2$ plot for $s_0=40\,GeV^2$ and $q^2=0\,GeV^2$. (a) For $f_1$ form factor of $1^3 P_0$ charmonium. (b) For $f_{V_+}$ form factor of $1^3 P_1$ charmonium.}
  	\label{fig:LTvsST}
\end{figure}

Another ingredient to obtain 
the numerical values of the right hand side of \cref{eq:eqsr} are the Borel parameter $M^2$ and the continuum threshold, $s_0$.
The Borel parameter, $M^2$, is an auxiliary parameter and physical quantities like the form factor should be independent of its value. 
Due to the approximations used in the method, a residual $M^2$ dependence might remain. 
To reduce this dependence, a suitable region of $M^2$ should be chosen such that the predictions are independent of this parameter
\cite{Navarra:2000ji,Bracco:2001dj,Matheus:2002nq,Navarra:2001ju,Bracco:2004rx,Carvalho:2005et,Matheus:2005yu,Bracco:2007sg,Rodrigues:2010ed,Bracco:2011pg}. 
On the other hand, the threshold parameter, $s_0$, is mainly related to hadron mass in question. It is usually chosen as $s_0\simeq(m_{hadron}\pm0.5\, GeV)^2$. 
The upper limit of the $M^2$ region can be determined by requiring the contribution of the lowest pole to be at least 50\%, whereas the lower limit can be determined by requiring the convergence of the twist expansion.
%
In \cref{fig:LTvsST}, the relative contribution of the subleading twist distribution amplitudes to the final result for the form factors $f_1$ and $f_{V_+}$ are shown as two examples. As can be seen in the figure, the relative contribution of the subleading twist distribution amplitudes diminish rapidly with increasing $M^2$ for the form factor $f_1$, but for $f_{V_+}$ the reduction rate is smaller. For $M^2 > 5~GeV^2$, the contribution of the subleading twist terms to all the form factors are found to be less than 15\%. Note that this is true, even though the subleading DAs are enhanced by the heavy meson mass.
 
To analyse the pole contribution, the following ratio is studied \cite{Bracco:2011pg}:
\begin{equation}
Pole=\dfrac{\int_{0}^{s_0} ds\,\rho^{QCD}(s) e^{-\frac{s}{M^2}}}{\int_{0}^{\infty} ds\,\rho^{QCD}(s) e^{-\frac{s}{M^2}}}.
\label{eq:pole}
\end{equation}

\begin{figure}\centering	
	    \includegraphics[width=0.45\textwidth]{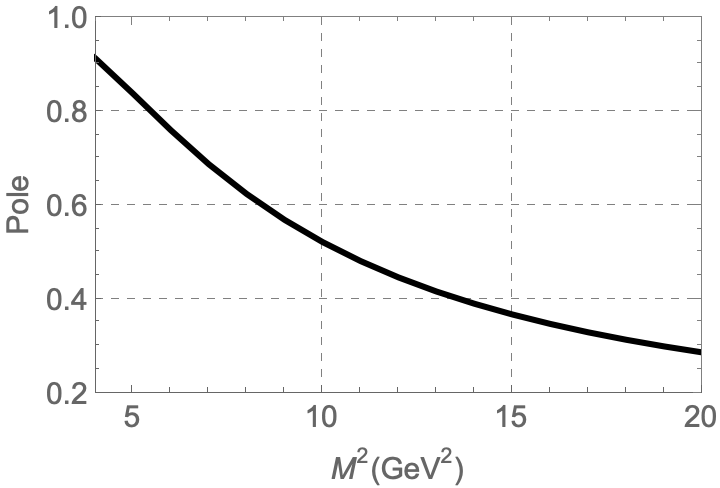}
  	\caption{$Pole$-$M^2$ plot of $1^3 P_2$ states's $f_{T_0}$ form factor for $q^2=0\, GeV^2$ and $s_0=40 \, GeV^2$.}
 	\label{fig:polecont}
\end{figure}

\begin{figure}
  	\subfloat[]{
 		\includegraphics[width=0.45\textwidth]{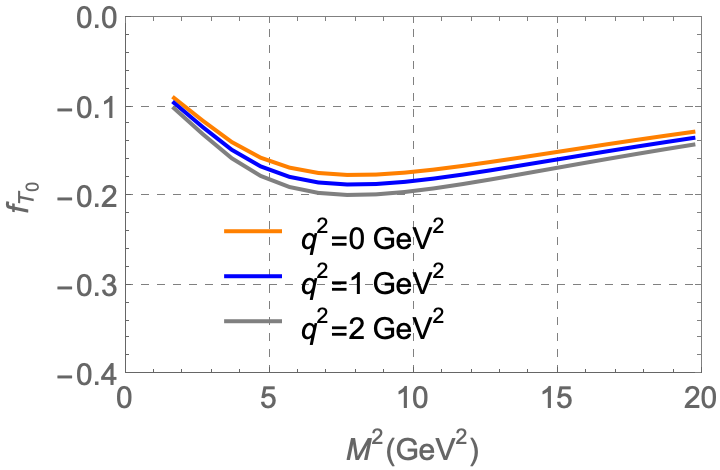}}
	\subfloat[]{
  		\includegraphics[width=0.45\textwidth]{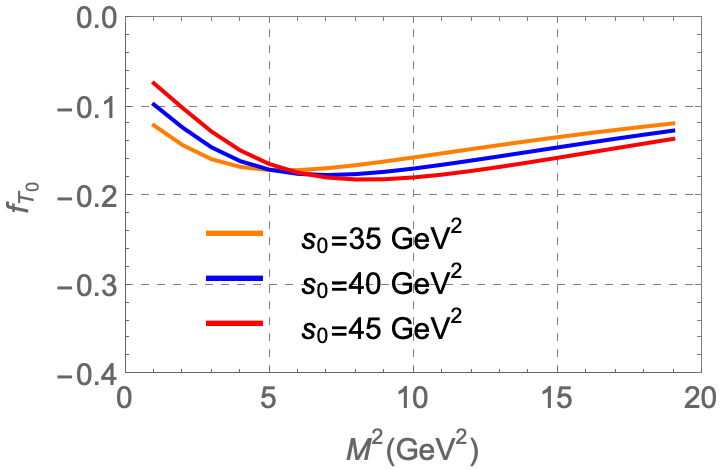}}\\
  	\caption{(a) $1 ^3P_2$ state's $f_{T_0}$ form factor with respect to $M^2$ for $s_0=40\, GeV^2$ and $q^2=0$ $GeV^2$ (blue line), 1 $GeV^2$ (orange line), and 2 $GeV^2$ (gray line). (b) $1 ^3P_2$ state's $f_{T_0}$ form factor with respect to $M^2$ for $q^2=0\,GeV^2$ and $s_0=35$ $GeV^2$ (orange line), 40 $GeV^2$ (blue line), and 45 $GeV^2$ (red line).}
  	\label{fig:fig6}	
\end{figure}

The pole contribution to the form factor is shown in \cref{fig:polecont}. As can be seen from this figure, the pole contribution is more than 50\% if $M^2<10~GeV^2$. It is found the this is also correct for all the form factors. Note that, it analysis, once the Borel region is determined, it is expected that the physical parameters are independent of the Borel parameter within this region. As an example, in \cref{fig:fig6}, the $M^2$ dependence of the form factor $f_{T_0}(q^2)$ for $1^3P_2$ state is shown for various $s_0$ and $q^2$ values. It is seen in this figure that for the chosen $s_0$ values, the prediction on the form factor is practically independent of the value of $M^2$ within the determined region.
The same conclusion also holds for other form factors. Hence, in further analysis, all the form factors will be analysed in the Borel region $5~GeV^2 < M^2 < 10~GeV^2$ for the $s_0$ values $s_0=40\pm5~GeV^2$.


\begin{figure}\centering	
	    \includegraphics[width=0.45\textwidth]{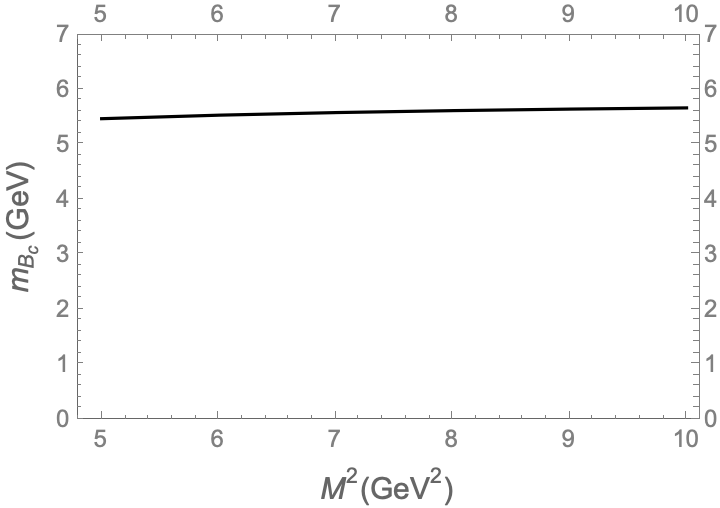}
  	\caption{$m_{B_c}$-$M^2$ plot (where $m_{B_c}$ is computed from $1 ^3 P_0$ state's $f_1$ form factor) for $q^2=0\, GeV^2$ and $s_0=40\, GeV^2$.}
 	\label{fig:mBcforf1}
\end{figure}

\begin{figure}\centering
  		\includegraphics[width=0.5\textwidth]{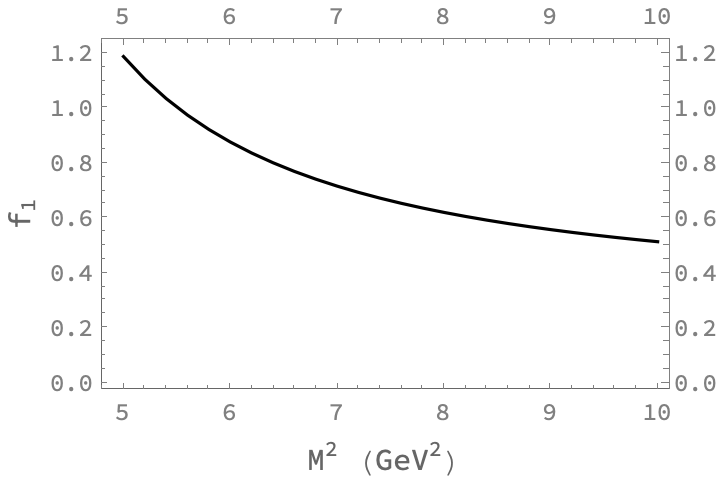}
  	\caption{$1 ^3P_0$ state's $f_1$ form factor with respect to $M^2$ for experimentally measured $B_c$ mass ($6.28\, GeV$) for $q^2=0$ and $s_0=40~GeV^2$.}
  	\label{fig:expmBc}	
\end{figure}

Using the parameters discussed above,  numerical values of the right hand side of the sum rules given in \cref{eq:eqsr} can be obtained. 
To predict numerical values of the form factors, it is necessary to choose a value for the mass of the $B_c$ meson. 
Frequently, the value of the hadron in question is picked from the experiment, if it is experimentally known. But this assumes that \cref{eq:eqsr} predicts the mass of the hadron precisely. 
In order to avoid this assumption, the mass of the $B_c$ meson will also be obtained from \cref{eq:eqsr} using the relation
\begin{align}
    m_{B_c}^2 = \frac{\int_0^{s_0} ds s \rho^{QCD}(q^2,s) e^{-\frac{s}{M^2}}}{\int_0^{s_0} ds \rho^{QCD}(q^2,s) e^{-\frac{s}{M^2}}}
    \label{eq:eqsrmass}
\end{align}
The masses obtained using \cref{eq:eqsrmass} should be stable with respect to variations of $M^2$ in chosen region.
In \cref{fig:mBcforf1}, the dependence of $m_{B_c}$ obtained from \cref{eq:eqsrmass} on the Borel parameter $M^2$
is presented for $s_0=40~GeV^2$ using the sum rules for the form factor $f_1(0)$. As can be seen in this figure, the mass is practically independent of the borel parameter $M^2$ in the chosen region.
For comparison, in \cref{fig:expmBc}, the Borel parameter dependence of $f_1$ is shown when \cref{eq:eqsrmass} is used for $m_{B_c}^2$ and when its experimental value is used. It is clearly seen that the form factors has a stronger dependence on the Borel parameter when the experimental value is used. 
This shows that after the continuum and higher states are subtracted, the correlation function does have the form given in the left hand side of \cref{eq:eqsr} but the exponent predicted by the correlation function deviates from the experimentally observed mass. 
This is an expected behaviour as the obtained correlation function is only approximate and we strongly believe the correct procedure should be to use the mass of $B_c$ which is predicted by the calculated correlation function.

In  \cref{table:tab},  the working region of the Borel parameter and the $B_c$ mass obtained from the corresponding sum rules is shown for each of the form factors.
As can be noted from the table, the predicted masses for $B_c$ are in general lower than the experimentally observed value.
Note that this is not a consequence of the approach used in this work. 
When the sum rules for the form factors obtained in \cite{Azizi:2009ny} is analyzed using the approach presented in this work, the obtained masses are compatible with the ones presented in \cref{table:tab}.

\begin{table}
\begin{center}
\begin{tabular}{|c|c|c|c|}
\hline
     Charmonia & Form Factor &  State  & $m_{B_c}\, (GeV)$\\ \hline 
     \multirow{6}{*}{$^3P_0$}&\multirow{3}{*}{$f_1^n$} &$n=1$ & $5.67\pm0.02$  \\\cline{3-4}
     & & $n=2$ & $5.59\pm 0.01$  \\ \cline{3-4}
     & & $n=3$ & $5.38\pm0.01$  \\ \cline{2-4}
     & \multirow{3}{*}{$f_2^n$} &$n=1$ & $5.72\pm 0.01$  \\\cline{3-4}
     & & $n=2$ & $5.59\pm 0.02$   \\ \cline{3-4}
     & & $n=3$ & $5.37\pm 0.01$   \\ \hline
     \multirow{6}{*}{$^1P_1$}&\multirow{3}{*}{$f_V^n$} &$n=1$ & $5.88\pm0.05$  \\\cline{3-4}
     & & $n=2$ & $5.86\pm0.05$  \\ \cline{3-4}
     & & $n=3$ & $5.71\pm0.08$ \\ \cline{2-4}
     & \multirow{3}{*}{$f_{V_+}^n$} &$n=1$ & $5.45\pm 0.05$  \\\cline{3-4}
     & & $n=2$ & $5.54\pm 0.05$  \\ \cline{3-4}
     & & $n=3$ & $5.54\pm0.15$  \\\cline{2-4}
     & \multirow{3}{*}{$f_{V_-}^n$} &$n=1$ & $5.82\pm 0.05$  \\\cline{3-4}
     & & $n=2$ & $5.73\pm 0.05$  \\ \cline{3-4}
     & & $n=3$ & $5.59\pm0.15$  \\ \cline{2-4}
     & \multirow{3}{*}{$f_{V_0}^n$} &$n=1$ & $5.20\pm 0.05$  \\\cline{3-4}
     & & $n=2$ & $5.11\pm 0.04$  \\ \cline{3-4}
     & & $n=3$ & $5.10\pm0.04$  \\ \hline
     \multirow{6}{*}{$^3P_1$}&\multirow{3}{*}{$f_V^n$} &$n=1$ & $5.80\pm0.12$  \\\cline{3-4}
     & & $n=2$ & $5.72\pm 0.12$  \\ \cline{3-4}
     & & $n=3$ & $5.54\pm 0.12$  \\ \cline{2-4}
     & \multirow{3}{*}{$f_{V_+}^n$} &$n=1$ & $5.47\pm 0.15$  \\\cline{3-4}
     & & $n=2$ & $5.47\pm 0.12$  \\ \cline{3-4}
     & & $n=3$ & $5.37\pm0.12$  \\\cline{2-4}
     & \multirow{3}{*}{$f_{V_-}^n$} &$n=1$ & $5.28\pm 0.05$  \\\cline{3-4}
     & & $n=2$ & $5.51\pm 0.05$  \\ \cline{3-4}
     & & $n=3$ & $5.40\pm0.02$  \\ \cline{2-4}
     & \multirow{3}{*}{$f_{V_0}^n$} &$n=1$ & $4.45\pm 0.02$  \\\cline{3-4}
     & & $n=2$ & $4.05\pm 0.10$  \\ \cline{3-4}
     & & $n=3$ & $4.72\pm0.02$  \\ \hline
    \multirow{6}{*}{$^3P_2$}&\multirow{3}{*}{$f_T^n$} &$n=1$ & $5.22\pm 0.12$  \\\cline{3-4}
     & & $n=2$ & $5.19\pm 0.10$  \\ \cline{3-4}
     & & $n=3$ & $5.14\pm0.10$  \\ \cline{2-4}
     & \multirow{3}{*}{$f_{T_+}^n$} &$n=1$ & $4.58\pm 0.15$  \\\cline{3-4}
     & & $n=2$ & $4.60\pm 0.08$  \\ \cline{3-4}
     & & $n=3$ & $4.57\pm0.08$  \\\cline{2-4}
     & \multirow{3}{*}{$f_{T_-}^n$} &$n=1$ & $4.58\pm 0.15$  \\\cline{3-4}
     & & $n=2$ & $4.60\pm 0.08$  \\ \cline{3-4}
     & & $n=3$ & $4.57\pm0.08$  \\ \cline{2-4}
     & \multirow{3}{*}{$f_{T_0}^n$} &$n=1$ & $4.88\pm 0.09$  \\\cline{3-4}
     & & $n=2$ & $4.87\pm 0.09$  \\ \cline{3-4}
     & & $n=3$ & $4.24\pm0.07$  \\ \hline 
\end{tabular}
\end{center}
\caption{Mass prediction of $B_c$ meson from all form factors in $5~GeV^2\le M^2\le 10~GeV^2$. Here the error comes from $s_0$ region ($35~GeV^2\leq s_0 \leq 45~GeV^2$).}
      \label{table:tab}
\end{table}
In the semileptonic decays of $B_c$ mesons, the relevant $q^2$ values do not always lie in the region where the sum rules predictions are reliable. For this reason, the form factors obtained using QCD sum rules in the region of $q^2$ where the sum rules' predictions are reliable needs to be extrapolated to whole of
the physically relevant region.

\begin{figure}\centering
  		\includegraphics[width=0.45\textwidth]{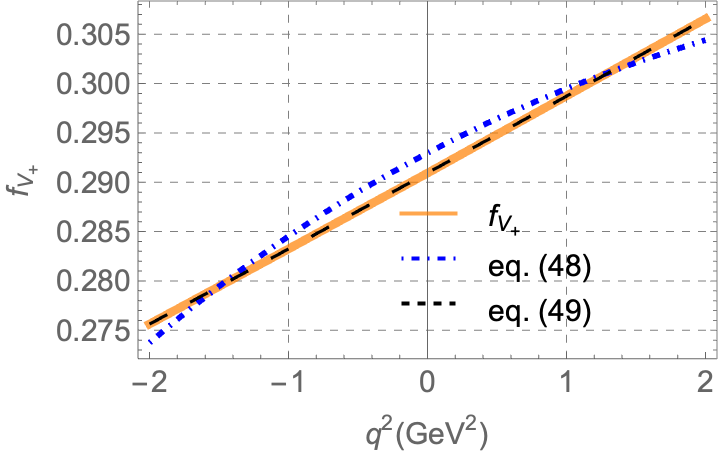}
  	\caption{$f_{V_+}$-$q^2$ plot of $1 ^1P_1$ axial-vector meson and its fit functions. }	
  	\label{fig:fig9}
\end{figure}

%
%
For the extrapolation, a suitable chosen extrapolation function is used. The explicit form of this function is another source of error in the predictions that depend the values of the form factors in the whole kinematical region. To estimate this additional error,
the following two commonly used functions are used to extrapolate the form factors to the whole kinematical $q^2$ region.
\begin{equation}
	f^n_i(q^2)=\dfrac{a}{\bigg(1-\dfrac{q^2}{m_{fit}^2}\bigg)}+\dfrac{b}{\bigg(1-\dfrac{q^2}{m_{fit}^2}\bigg)^2},
	\label{eq:eqfit}
\end{equation}
\begin{equation}
	f^n_i(q^2)=F_0\, exp\bigg\{a\bigg(\dfrac{q^2}{m_{fit}^2}\bigg)+b\bigg(\dfrac{q^2}{m_{fit}^2}\bigg)^2\bigg\},
	\label{eq:expfit}
\end{equation}
where $F_0$, $a$, $b$, and $m_{fit}$ are the fit parameters. These parameters are chosen so that this function overlaps with the form factor prediction obtained in the region of $q^2$ where the predictions are reliable. The obtained values are  shown in \cref{app:fpexp}. 
In \cref{eq:eqfit}, the fit parameter $m_{fit}$ obtained from the fit to various form factors all turned out to be very close to $m_{fit}=5.18~GeV$. Hence, we fixed this parameter at this value for the fits. 
In the region $-2~
GeV^2 < q^2 < 2~GeV^2$, the two fit functions reliably reproduce the obtained predictions. The largest difference between the best fits obtained for the two different fit functions is observed in $f_{V_+}$ for $1^1P_1$ charmonium. In \cref{fig:fig9} this form factors along with the fits obtained by the two functions are shown. As can be seen from this figure, even the largest difference between the two functions is less than 1\%. 



\section{Semileptonic Decay Rates}
\label{sec:decay}
Once the formfactors are obtained, they can be used to analyze the semileptonic decay rates of the $B_c$ meson into P-wave charmonia.
The decay rates can be expressed in terms of helicity amplitudes as \cite{Ivanov:2005fd, Ebert:2010zu,Issadykov:2017wlb,Faustov:2019mqr}:
\begin{equation}
	\dfrac{d\Gamma(B_c\rightarrow c\bar{c}l\bar{\nu})}{dq^2}=\dfrac{G_F^2}{(2\pi)^3}|V_{bc}|^2\dfrac{\lambda^{1/2}(q^2-m_l^2)^2}{24\,m_{B_c}^3\,q^2}\bigg[HH^\dagger\bigg(1+\dfrac{m_l^2}{2q^2}\bigg)+\dfrac{3m_l^2}{2q^2}H_tH_t^\dagger\bigg]
	\label{eq:eq37}
\end{equation}
where $\lambda=m_{B_c}^4+m_{c\bar{c}}^4+q^4-2(m_{B_c}^2 m_{c\bar{c}}^2+m_{c\bar{c}}^2 q^2+m_{B_c}^2 q^2)$, $m_{B_c}=6.28\, GeV$, $m_{c\bar{c}}$ is the corresponding charmonium mass ($m_S$, $m_{A}$, $m_T$) and $m_l$ is the corresponding lepton mass ($m_e=0.5\, MeV$, $m_\mu=105.66\, MeV$, $m_\tau=1.78\, GeV$) and
\begin{equation}
	HH^\dagger\equiv H_+H_+^\dagger+H_-H_-^\dagger+H_0H_0^\dagger
	\label{eq:eq38}
\end{equation}
where the subscripts $\pm$, 0 ,t denote transverse, longitudinal and time like helicity components, respectively and they are defined in terms of the form factors as

\begin{figure}	
	\subfloat[$1^3P_0$]{
  		\includegraphics[width=0.4\textwidth]{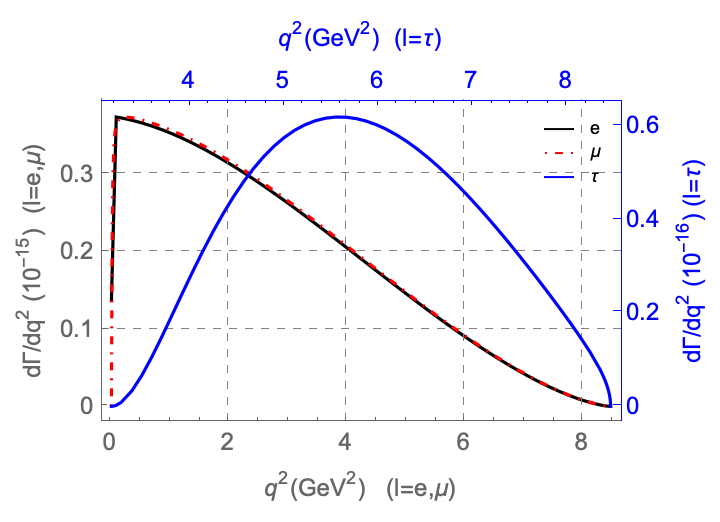}}
	\subfloat[$2^3P_0$]{
  		\includegraphics[width=0.4\textwidth]{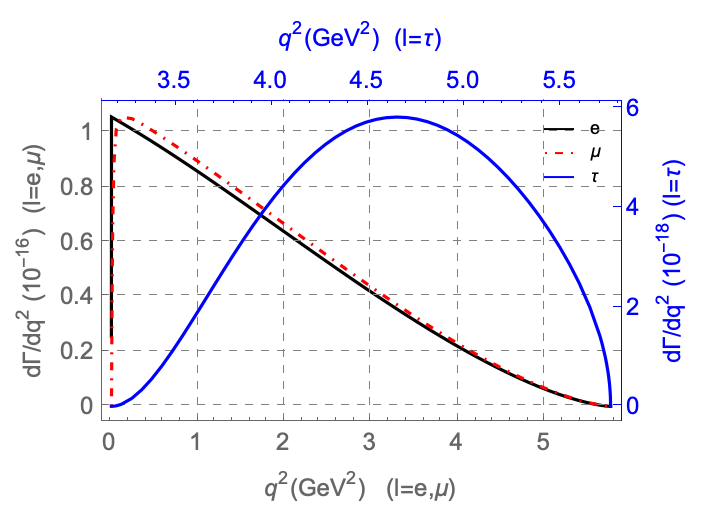}}\\
  	\centering{\subfloat[$3^3P_0$]{
  		\includegraphics[width=0.45\textwidth]{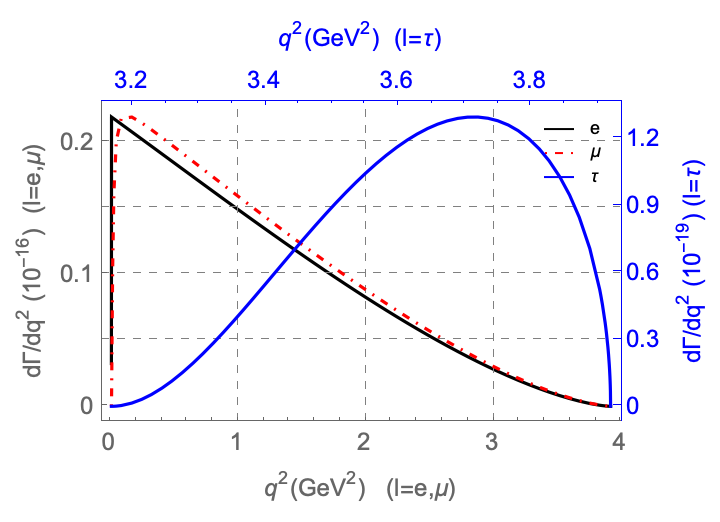}}}
  	\caption{$d\Gamma/dq^2$-$q^2$ plots for decays involving  $^3P_0$ scalar meson and its excited states for $e$, $\mu$, and $\tau$ leptons in the final states. There are two different scales for $q^2$ and $d\Gamma/dq^2$ in each frame since left and bottom axis of frame are for $e$ and $\mu$ leptons and right and up axis of frame are for $\tau$ lepton.}
  	\label{fig:fig12}
\end{figure}

\begin{figure}
	\subfloat[$1^3P_1$]{
  		\includegraphics[width=0.4\textwidth]{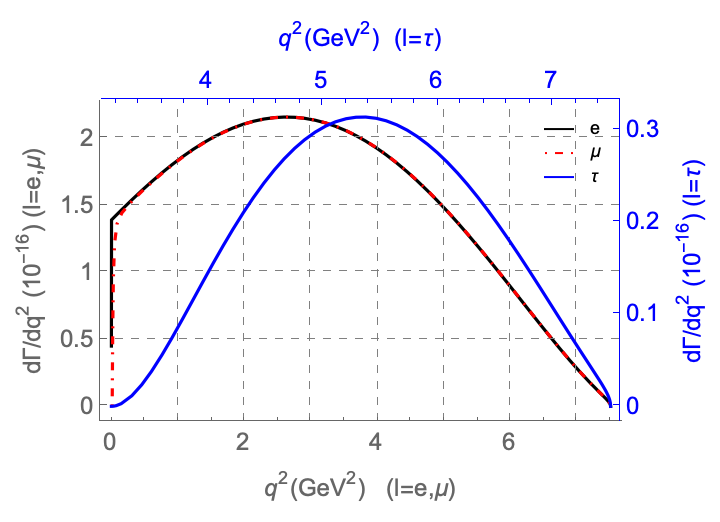}}
	\subfloat[$2^3P_1$]{
  		\includegraphics[width=0.4\textwidth]{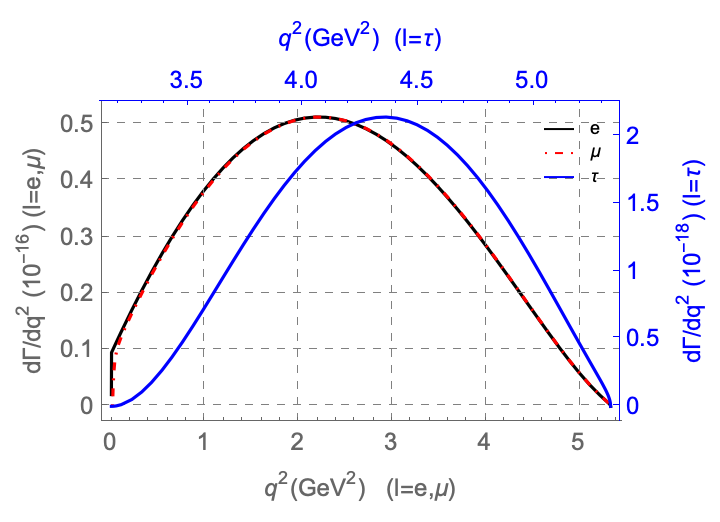}}\\
  	\centering{\subfloat[$3^3P_1$]{
  		\includegraphics[width=0.45\textwidth]{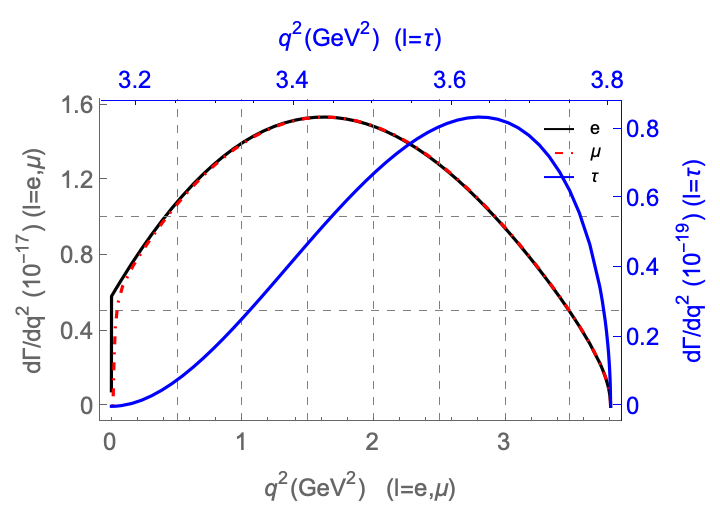}}}
  	\caption{$d\Gamma/dq^2$-$q^2$ plots for decays involving $^3P_1$ scalar meson and its excited states for $e$, $\mu$, and $\tau$ leptons}
  	\label{fig:fig13}
\end{figure}

\begin{figure}
	\subfloat[$1^1P_1$]{
  		\includegraphics[width=0.4\textwidth]{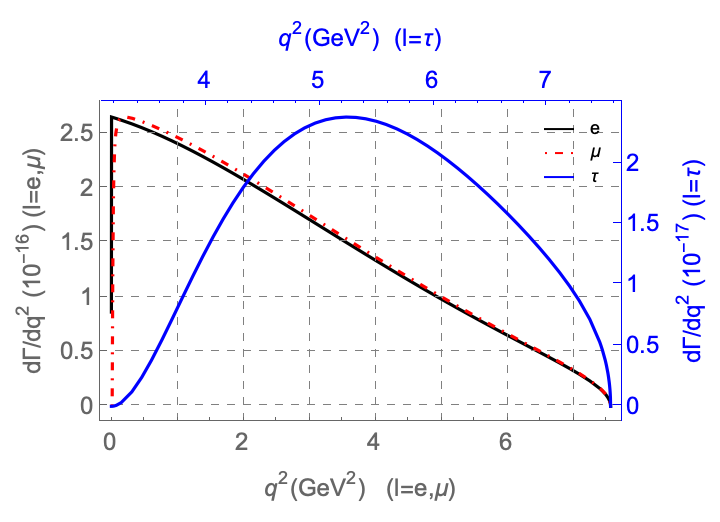}}
	\subfloat[$2^1P_1$]{
  		\includegraphics[width=0.4\textwidth]{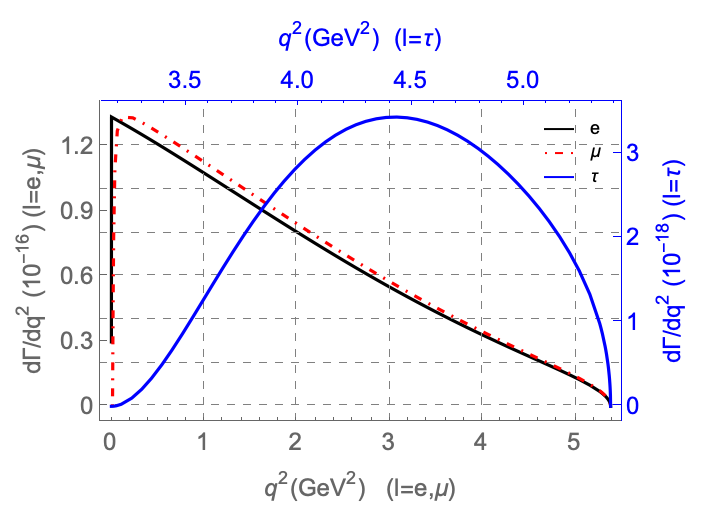}}\\
  	\centering{\subfloat[$3^1P_1$]{
  		\includegraphics[width=0.45\textwidth]{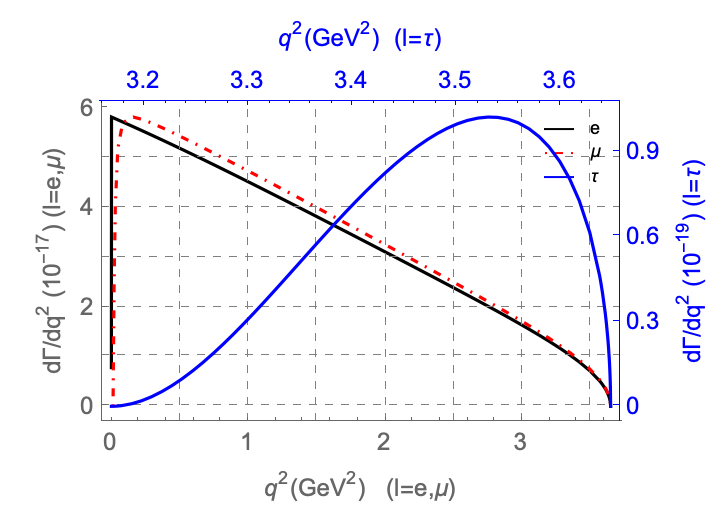}}}
  	\caption{$d\Gamma/dq^2$-$q^2$ plots for decays involving $^1P_1$ scalar meson and its excited states for $e$, $\mu$, and $\tau$ leptons}
  	\label{fig:fig14}
\end{figure}

\begin{figure}
	\subfloat[$1^3P_2$]{
  		\includegraphics[width=0.4\textwidth]{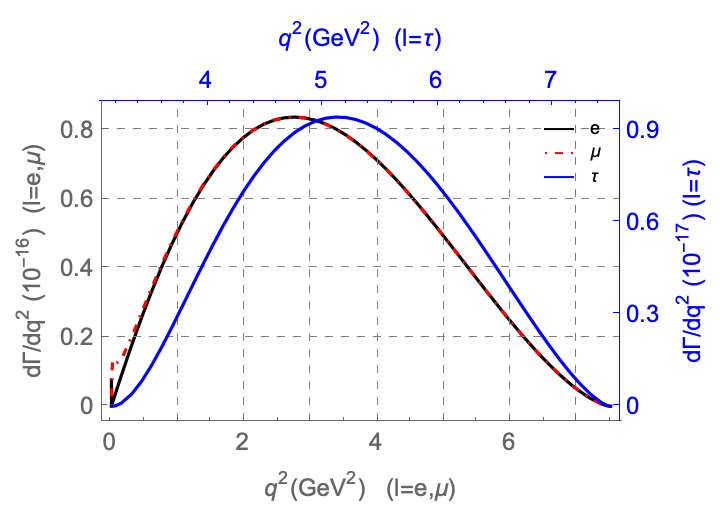}}
	\subfloat[$2^3P_2$]{
  		\includegraphics[width=0.4\textwidth]{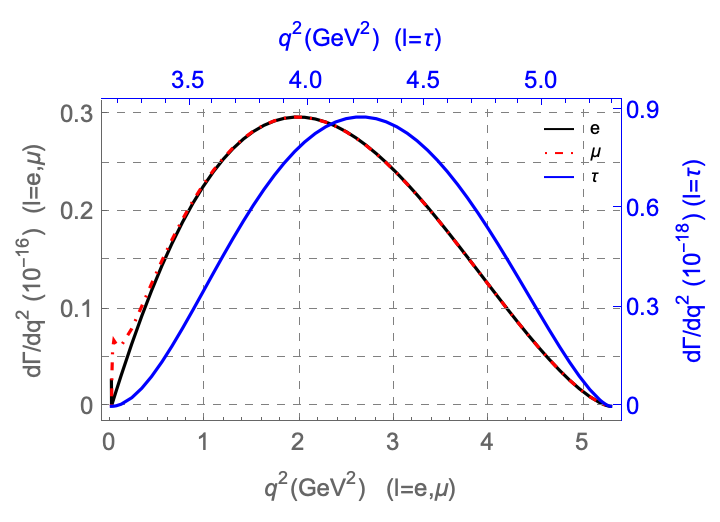}}\\
  	\centering{\subfloat[$3^3P_2$]{
  		\includegraphics[width=0.45\textwidth]{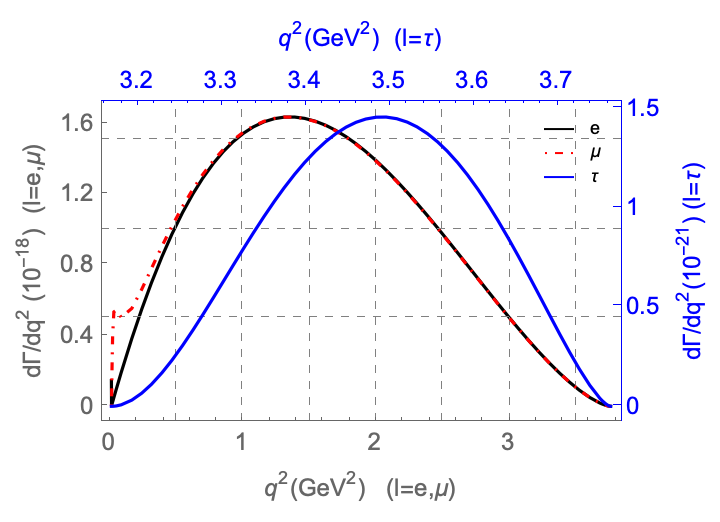}}}
  	\caption{$d\Gamma/dq^2$-$q^2$ plots for decays involving $^3P_2 $ scalar meson and its excited states for $e$, $\mu$, and $\tau$ leptons}	
  	\label{fig:fig15}
\end{figure}

\begin{itemize}
  \item Spin 0: $B_c \rightarrow S(^3 P_0)$ transition
  \begin{align}
	H_\pm &=0, \\
	H_0 &=\dfrac{\lambda^{1/2}}{\sqrt{q^2}}f^n_1(q^2),\nonumber \\
	H_t &=\dfrac{1}{\sqrt{q^2}}[(m_{B_c}^2-m_S^2)f^n_1(q^2)+q^2f^n_2(q^2)]\nonumber
	\label{eq:eq39}
\end{align}
where $m_S$ is the scalar meson mass.
 \item Spin 1: $B_c \rightarrow A(^3 P_1, ^1 P_1)$ transition
 \begin{align}
	H_\pm &=-(m_{B_c}+m_A)f^n_{V_0}(q^2)\mp\dfrac{\lambda^{1/2}}{m_{B_c}+m_A}f^n_V(q^2), \\
	H_0 &=\dfrac{1}{2m_A \sqrt{q^2}}\bigg\{-(m_{B_c}^2-m_{A}^2-q^2)(m_{B_c}+m_A)f^n_{V_0}(q^2)+\dfrac{\lambda^{1/2}}{m_{B_c}+m_A}\, f^n_{V_+}(q^2)\bigg\},\nonumber\\
	H_t &=\dfrac{\lambda^{1/2}}{2m_{A}\sqrt{q^2}}\bigg\{-(m_{B_c}+m_{A})f^n_{V_0}(q^2)+(m_{B_c}-m_A)f^n_{V_+}(q^2)+\dfrac{q^2}{m_{B_c}+m_A}\,f^n_{V_-}(q^2)\bigg\}.\nonumber
	\label{eq:eq40}
\end{align}
Notice that $f^n_{1_V}$, $f^n_{1_{V_+}}$, $f^n_{1_{V_-}}$, and $f^n_{1_{V_0}}$ form factors for the $^1P_1$ charmonium state and $f^n_{3_V}$, $f^n_{3_{V_+}}$, $f^n_{3_{V_-}}$, and $f^n_{3_{V_0}}$ form factors for the $^3P_1$ charmonium state are simplified as $f^n_{V}$, $f^n_{V_+}$, $f^n_{V_-}$, and $f^n_{V_0}$.
 \item Spin 2: $B_c \rightarrow T(^3 P_2)$ transition
 \begin{align}
	H_\pm &=\dfrac{\lambda^{1/2}}{2\sqrt{2}m_{B_c}m_T}\bigg\{(m_{B_c}+m_T)f^n_{T_0}(q^2)\pm \dfrac{\lambda^{1/2}}{m_{B_c}+m_T}f^n_T(q^2)\bigg\},\\
	H_0 &=\dfrac{\lambda^{1/2}}{2\sqrt{6}m_{B_c}m_T}\bigg\{(m_{B_c}+m_T)(m_{B_c}^2-m_T^2-q^2)f^n_{T_0}(q^2)-\dfrac{\lambda}{m_{B_c}}f^n_{T_+}(q^2)\bigg\},\nonumber\\
	H_t &=\sqrt{\dfrac{2}{3}}\dfrac{\lambda}{4m_{B_c}m_T^2\sqrt{q^2}}\bigg\{(m_{B_c}+m_T)f^n_{T_0}(q^2)-\dfrac{m_{B_c}^2-m_T^2}{m_{B_c}}f^n_{T_+}(q^2)-\dfrac{q^2}{m_{B_c}}f^n_{T_-}(q^2)\bigg\}.\nonumber
	\label{eq:eq41}
\end{align}
\end{itemize}
where $m_S$, $m_{AV}$, and $m_T$ are the corresponding charmonium mass and they are given in \cref{table:masses}.
The physical region for $q^2$ is defined as $m_l^2\leq q^2\leq(m_{B_c}-m_{c\bar{c}})^2$. 
In \cref{fig:fig12,fig:fig13,fig:fig14,fig:fig15} the differential $B_c$ decay rates are shown for various final states. In these figures, the fit function in \cref{eq:eqfit} is used. As can be seen from the figures, the differential decay rates for decays involving $e$ and $\mu$ are almost identical, whereas the differential decay rate for decays involving $\tau$ is significantly lower due to $\tau$'s heavy mass.

Finally, we present branching ratios for each ground state and excited states of P-wave charmonia and compare them with the results in the literature in \cref{table:tab5}. As can be seen from the table, the choice of the fit function does not influence the branching ratios for decays into scalar and tensor charmonia, whereas the branching ratios for the decays into axial vector charmonia change by  upto a factor two.
As is mentioned in the introduction, there are no QCDSR predictions for decays into final states containing excited  P-wave charmonia.
The presented results from literature are obtained using non-relativistic quark model computations.
As can be seen from the tables, our predictions for decays into final states $1 ^3P_0$ and $1 ^1P_1$ are within the range of other predictions
in the literature, where as for the final state $1 ^3P_2$, our predictions are slightly smaller, and for the final state $1 ^3P_1$ our prediction is slightly
larger than the existing predictions in the literature. As our predictions for decays not involving radially excited charmonia are compatible with the existing
results in the literature, this can be taken as an indication that the results for the excited state are also reliable.
It should also be noted that, keeping only the form factors $f_1$, $f_{V_0}$ and $f_{T_0}$ and setting all the other form factors to zero, the changes in
the branching ratios are less than $10\%$.


\begin{table}\centering
\footnotesize
\begin{tabular}{|*{1}{*{11}{l}}|}
\hline
& This & \cite{Azizi:2009ny} &\cite{Ebert:2010zu}&\cite{Ivanov:2006ni}&\cite{Hernandez:2006gt}& This & \cite{Azizi:2009ny} &  \cite{Ebert:2010zu}&\cite{Ivanov:2006ni}&\cite{Hernandez:2006gt}\\$\quad c\bar{c}$ & paper  & QCDSR & RQMQP & RQM & NRQM & paper & QCDSR & RQMQP & RQM &NRQM \\
& $l=e,\,\mu$ & $l=e$ & $l=e$ &$l=e$ &$l=e$ &  $l=\tau$ &  $l=\tau$ &  $l=\tau$ &$l=\tau$ &$l=\tau$  \\
\hline
$1^3 P_0$ & $0.114$ &$0.182$ & $0.087$ & $0.180$ & $0.110$ &  $0.0014$ & $0.018$ & $0.0075$ & $0.018$ & $0.013$  \\

 & $(0.119)$ & &  &  &  & $(0.0017)$  &  &  &  &   \\

$2^3 P_0$  &$0.020$ &$-$ & $0.014$ & $-$ &$-$ &  $6.8\times 10^{-4}$ & $-$ &  $6.3\times 10^{-4}$ &$-$ &$-$ \\

& $(0.019)$ & &  &  &  &  $(6.0\times 10^{-4})$ &  &  &  &   \\

$3^3P_0$  & $0.0027$ & $-$ & $-$ &$-$ &$-$ & $4.04\times 10^{-6}$ &  $-$ &  $-$ &$-$ &$-$ \\

& $(0.0024)$ & &  &  &  &  $(2.73\times 10^{-6})$ &  &  &  &   \\

$1^1 P_1$  & $0.077$ & $0.142$ & $0.096$ &$0.310$&$0.170$ & $0.0048$ & $0.0137$ &  $0.0077$&$0.027$ &$0.015$  \\

& $(0.108)$ & &  &  &  &  $(0.0104)$ &  &  &  &   \\

$2^1 P_1$  & $0.026$ & $-$ & $0.0021$ &$-$ &$-$ &  $3.5\times 10^{-4}$ & $-$ &  $0.00011$ &$-$ &$-$  \\

& $(0.025)$ & &  &  &  &  $(5.6\times 10^{-4})$ &  &  &  &   \\

$3^1P_1$ & $0.0089$ & $-$ & $-$ &$-$ &$-$ &  $1.97\times 10^{-6}$ &  $-$ &  $-$&$-$ &$-$ \\

& $(0.0124)$ & &  &  &  &  $(2.51\times 10^{-6})$ &  &  &  &   \\

$1^3 P_1$ & $0.080$ & $0.146$ & $0.082$ & $0.098$& $0.066$ & $0.0057$ & $0.0147$ &  $0.0092$ &$0.012$ & $0.0072$  \\

& $(0.112)$ & &  &  &  &  $(0.0079)$ &  &  &  &   \\

$2^3 P_1$ & $0.013$ & $-$ & $0.0085$ & $-$ &$-$ &  $1.9\times 10^{-4}$ & $-$ &  $3.9\times 10^{-4}$&$-$ & $-$ \\

& $(0.022)$ & &  &  &  &  $(2.51\times 10^{-4})$ &  &  &  &   \\

$3^3P_1$ & $0.003$ & $-$ & $-$ &$-$ &$-$ &  $2.16\times 10^{-6}$ &  $-$ &  $-$&$-$ &$-$ \\

& $(0.004)$ & &  &  &  &  $(1.34\times 10^{-6})$ &  &  &  &   \\

$1^3 P_2$ & $0.026$ & $-$ & $0.160$ & $0.200 $ & $0.130$ &   $0.0019$ &  $-$ &  $0.0093$& $0.014$ & $0.0093$ \\

& $(0.025)$ & &  &  &  &  $(0.0018)$ &  &  &  &   \\

$2^3 P_2$ & $0.0066$ & $-$ & $0.0033$ &$-$ &$-$ & $7.36\times 10^{-5}$ &  $-$ & $1.3\times 10^{-4}$&$-$ &$-$  \\

& $(0.0061)$ & &  &  &  &  $(7.0\times 10^{-5})$ &  &  &  &   \\

$3^3P_2$ & $0.00023$ & $-$ &  $-$ & $-$ &$-$ & $3.39\times 10^{-8}$ &  $-$ &  $-$&$-$ &$-$ \\

& $(0.00023)$ & &  &  &  &  $(2.47\times 10^{-8})$ &  &  &  &   \\
\hline
\end{tabular}	
\caption{Branching Ratios (in $\%$) of semi-leptonic $B_c$ decays to P-wave charmonia ($B_c\rightarrow (c\bar{c})\, l\, \bar{\nu}$) for fixed value of the $B_c$ lifetime $\tau_{B_c}=0.46\, ps$. In the column showing our results the numbers in paranthesis are the results obtained  for the exponential fit.}
\label{table:tab5}	
\end{table}

\section{Conclusions}
\label{sec:conc}
 
In this work, the form factors and differential decay rates of semileptonic decays of $B_c$ meson into P-wave charmonia and their excited states in the framework of LCSR. The excited states are parameterized by distribution amplitudes obtained from 
non-relativistic QM wave functions in \cite{Olpak:2016wkf}. 
In the literature, there are studies of $B_c$ decays into the P-wave charmonia in the ground states. For these states, the results obtained in
this work and those available in the literature are comparable. This indicates that the approach used in this work of using QM wave functions to model the charmonium states are reliable. Hence, it is reasonable to assume that 
the obtained results for the excited states are also reliable. 


\medskip

\section*{Acknowledgments}

 This research has been supported by TUBITAK (The Scientific and Technological Research Council of Turkey) under the grant no 117F090. The authors would like to thank M. A. Olpak for his useful discussions.

\medskip

\begin{appendix}

\section{Formulae for form factors}
\label{app:fff}

The analytic results of the form factors are below
\begin{align}
	f^n_1(q^2)  \kappa_{f^n_1} e^{-\frac{m_{B_c}^2}{M^2}}&= \frac{3}{2} m_b f_S m_S^2 e^{-\frac{s_0}{M^2}} \int _0^1 du\int _0^u dt\frac{\phi _S(u)
   \delta \left(s_0-s\left(t,q^2,m_S^2\right)\right)}{t}\\\nonumber&+\frac{3 m_b f_S m_S^2}{2 M^2}
   \int _0^{s_0} ds\int _0^1 du\int _0^u dt\frac{e^{-\frac{s}{M^2}} \phi _S(u) \delta
   \left(s-s\left(t,q^2,m_S^2\right)\right)}{t}\\\nonumber&+\frac{3}{4} m_b f_S
   m_S^2 e^{-\frac{s_0}{M^2}} \int _0^1 du\int _0^u dt\int _0^u dv\frac{\phi _S(v) \delta
   \left(s_0-s\left(t,q^2,m_S^2\right)\right)}{t (v-1)}\\\nonumber&-\frac{3}{4} m_b f_S
   m_S^2 e^{-\frac{s_0}{M^2}} \int _0^1 du\int _0^u dt\int _u^1 dv\frac{\phi _S(v) \delta
   \left(s_0-s\left(t,q^2,m_S^2\right)\right)}{t v}\\\nonumber&+\frac{3 m_b f_S m_S^2}{4
   M^2}
   \int _0^{s_0} ds\int _0^1 du\int _0^u dt\int _0^u dv\frac{e^{-\frac{s}{M^2}} \phi _S(v)
   \delta \left(s-s\left(t,q^2,m_S^2\right)\right)}{t (v-1)}\\\nonumber&-\frac{3 m_b f_S m_S^2}{4 M^2} \int _0^{s_0} ds\int _0^1 du\int _0^u dt\int
   _u^1 dv\frac{e^{-\frac{s}{M^2}} \phi _S(v) \delta
   \left(s-s\left(t,q^2,m_S^2\right)\right)}{t v}\\\nonumber&-\frac{3}{2} m_b
   f_S \int _0^{s_0} ds\int _0^1 du\frac{e^{-\frac{s}{M^2}} \phi _S(u) \delta
   \left(s-s\left(u,q^2,m_S^2\right)\right)}{u}
 \end{align}
where 
\begin{align}
    s(u,q^2,m^2) = \bar{u} m^2 + \frac{m_b^2- \bar{u}q^2}{u} 
\end{align}

\begin{align}
	f^n_2(q^2)  \kappa_{f^n_2} e^{-\frac{m_{B_c}^2}{M^2}}&= -\frac{3}{2} m_b f_S m_S^2 e^{-\frac{s_0}{M^2}} \int _0^1 du\int _0^u dt\frac{(t-2) \phi
   _S(u) \delta \left(s_0-s\left(t,q^2,m_S^2\right)\right)}{t^2}\\\nonumber&-\frac{3 m_b
   f_S m_S^2}{2
   M^2} \int _0^{s_0} ds\int _0^1 du\int _0^u dt\frac{(t-2) e^{-\frac{s}{M^2}} \phi
   _S(u) \delta \left(s-s\left(t,q^2,m_S^2\right)\right)}{t^2}\\\nonumber&-\frac{3}{4} m_b f_S m_S^2 e^{-\frac{s_0}{M^2}} \int _0^1 du\int _0^u dt\int
   _0^u dv\frac{(t-2) \phi _S(v) \delta \left(s_0-s\left(t,q^2,m_S^2\right)\right)}{t^2
   (v-1)}\\\nonumber&+\frac{3}{4} m_b f_S m_S^2 e^{-\frac{s_0}{M^2}} \int _0^1 du\int
   _0^u dt\int _u^1 dv\frac{(t-2) \phi _S(v) \delta
   \left(s_0-s\left(t,q^2,m_S^2\right)\right)}{t^2 v}\\\nonumber&-\frac{3 m_b f_S m_S^2}{4
   M^2}
   \int _0^{s_0} ds\int _0^1 du\int _0^u dt\int _0^u dv\frac{(t-2) e^{-\frac{s}{M^2}} \phi
   _S(v) \delta \left(s-s\left(t,q^2,m_S^2\right)\right)}{t^2 (v-1)}\\\nonumber&+\frac{3 m_b f_S m_S^2 }{4 M^2}\int _0^{s_0} ds\int _0^1 du\int _0^u dt\int _u^1 dv\frac{(t-2)
   e^{-\frac{s}{M^2}} \phi _S(v) \delta \left(s-s\left(t,q^2,m_S^2\right)\right)}{t^2
   v}\\\nonumber&+\frac{3}{2} m_b f_S \int _0^{s_0} ds\int
   _0^1 du\frac{e^{-\frac{s}{M^2}} \phi _S(u) \delta
   \left(s-s\left(u,q^2,m_S^2\right)\right)}{u}
 \end{align}

\begin{align}
	f^n_V(q^2) \kappa_{f^n_V} e^{-\frac{m_{B_c}^2}{M^2}}&= 3 f_{A_\perp} \int _0^{s_0}ds \int _0^1 du\frac{e^{-\frac{s}{M^2}} \phi _{A_\perp}(u)
   \delta \left(s-s\left(u,q^2,m_A^2\right)\right)}{u}
   \label{eq:eq44}
\end{align}

\begin{align}
	f^n_{V_0}(q^2)\kappa_{f^n_{V_0}} e^{-\frac{m_{B_c}^2}{M^2}}&=-\frac{3 m_A^2}{2 M^2} \int _0^{s_0}ds\int _0^1 du\int _0^u dt\frac{e^{-\frac{s}{M^2}} \delta
   \left(s-s\left(t,q^2,m_A^2\right)\right)}{t^2}\\\nonumber&\qquad\qquad\times \left(2 m_A f_{A_\parallel} m_b (u-t) \phi
   _{A_\parallel}(u)+f_{A_\perp} \phi _{A_\perp}(u) \left(2 m_b^2+M^2
   t\right)\right)\\\nonumber&+3 m_A^2 m_b e^{-\frac{s_0}{M^2}} \int
   _0^1 du\int _0^u dt\frac{\delta \left(s_0-s\left(t,q^2,m_A^2\right)\right)}{t^2}\\\nonumber&\qquad\qquad \times \left(m_A
   f_{A_\parallel} (t-u) \phi _{A_\parallel}(u)-f_{A_\perp} m_b \phi
   _{A_\perp}(u)\right)\\\nonumber&+\frac{3 m_A^2}{2
   M^2} \int _0^{s_0} ds\int _0^1 du\int
   _0^u dt\int _0^u dv\frac{e^{-\frac{s}{M^2}} \delta \left(s-s\left(t,q^2,m_A^2\right)\right)}{t^2 (v-1)}\\\nonumber&\qquad\qquad \times
   \left(m_A f_{A_\parallel} m_b (t-u) \phi _{A_\parallel}(v)+(2 u-1) f_{A_\perp} \phi
   _{A_\perp}(v) \left(2 m_b^2+M^2 t\right)\right)\\\nonumber&-\frac{3 m_A^2}{2
   M^2} \int _0^{s_0} ds\int _u^1 du\int _0^1 dt\int
   _0^u dv\frac{e^{-\frac{s}{M^2}} \delta \left(s-s\left(t,q^2,m_A^2\right)\right) }{t^2 v}\\\nonumber&\qquad\qquad \times \left(m_A
   f_{A_\parallel} m_b (t-u) \phi _{A_\parallel}(v)+(2 u-1) f_{A_\perp} \phi
   _{A_\perp}(v) \left(2 m_b^2+M^2 t\right)\right)\\\nonumber&+\frac{3}{2} m_A^2 m_b e^{-\frac{s_0}{M^2}}\int _0^1 du\int _0^u dt\int
   _u^1 dv\frac{\delta \left(s_0-s\left(t,q^2,m_A^2\right)\right)}{t^2 v}\\\nonumber&\qquad\qquad \times \left(m_A f_{A_\parallel}
   (u-t) \phi _{A_\parallel}(v)+2 (1-2 u) f_{A_\perp} m_b \phi
   _{A_\perp}(v)\right)\\\nonumber&+\frac{3}{2} m_A^2 m_b e^{-\frac{s_0}{M^2}}
   \int _0^1 du\int _0^u dt\int _0^u dv\frac{\delta
   \left(s_0-s\left(t,q^2,m_A^2\right)\right)}{t^2
   (v-1)}\\\nonumber&\qquad\qquad \times \left(m_A f_{A_\parallel} (t-u) \phi
   _{A_\parallel}(v)+2 (2 u-1) f_{A_\perp} m_b \phi _{A_\perp}(v)\right)\\\nonumber&+\frac{3}{2} m_A f_{A_\parallel} m_b \int _0^{s_0}ds\int _0^1 du\int
   _u^1 dv\frac{e^{-\frac{s}{M^2}} \phi _{A_\parallel}(v) \delta
   \left(s-s\left(u,q^2,m_A^2\right)\right)}{u v}\\\nonumber&+3 m_A f_{A_\parallel} m_b
   \int _0^{s_0} ds\int _0^1 du\int _0^u dv\frac{e^{-\frac{s}{M^2}} \phi _{A_\parallel}(v)
   \delta \left(s-s\left(u,q^2,m_A^2\right)\right)}{2 u-2 u v}\\\nonumber&+\frac{3}{2}
   f_{A_\perp} \int _0^{s_0} ds\int _0^1 du\frac{e^{-\frac{s}{M^2}} \phi
   _{A_\perp}(u) \left(u^2 m_A^2+m_b^2+q^2\right) \delta
   \left(s-s\left(u,q^2,m_A^2\right)\right)}{u^2}
   \label{eq:eq45}
\end{align}

\begin{align}
	f^n_{V_+}(q^2)\kappa_{f^n_{V_+}} e^{-\frac{m_{B_c}^2}{M^2}}&=\frac{3f_{A_\perp}}{2} \int _0^{s_0} ds\int _0^1 du\frac{e^{-\frac{s}{M^2}} \delta
   \left(s-s\left(u,q^2,m_A^2\right)\right) \phi _{A_\perp}(u)}{u}
   \\\nonumber&+\frac{3m_A e^{-\frac{s_0}{M^2}}}{2 M^2} \int _0^1 du\int _0^u dt\frac{\delta
   \left(s_0-s\left(t,q^2,m_A^2\right)\right) }{t^2}\\\nonumber&\qquad\qquad \times \left(t f_{A_\perp} m_A \phi
   _{A_\perp}(u) M^2+2 f_{A_\parallel} \left(M^2+(t-u) m_A^2\right) m_b \phi
   _{A_\parallel}(u)\right)\\\nonumber&+\frac{3m_A}{2 M^4} \int _0^{s_0} ds\int
   _0^1 du\int _0^u dt\frac{e^{-\frac{s}{M^2}} \delta \left(s-s\left(t,q^2,m_A^2\right)\right)}{t^2}\\\nonumber&\qquad\qquad \times
   \left(t f_{A_\perp} m_A \phi _{A_\perp}(u) M^2+2 f_{A_\parallel} \left(M^2+(t-u)
   m_A^2\right) m_b \phi _{A_\parallel}(u)\right)\\\nonumber&+\frac{3
   e^{-\frac{s_0}{M^2}}m_A}{2 M^2} \int _0^1 du\int _0^u dt\int _0^u dv\frac{\delta
   \left(s_0-s\left(t,q^2,m_A^2\right)\right)}{t^2 (v-1)} \\\nonumber&\qquad\qquad \times\left(t (1-2 u) f_{A_\perp} m_A \phi
   _{A_\perp}(v) M^2+f_{A_\parallel} \left(M^2+(t-u) m_A^2\right) m_b \phi
   _{A_\parallel}(v)\right) \\\nonumber&-\frac{3 
   e^{-\frac{s_0}{M^2}} m_A}{2 M^2} \int _0^1 du\int _0^u dt\int _u^1 dv\frac{\delta
   \left(s_0-s\left(t,q^2,m_A^2\right)\right) }{ t^2 v}\\\nonumber&\qquad\qquad \times\left(t (1-2 u) f_{A_\perp} m_A \phi
   _{A_\perp}(v) M^2+f_{A_\parallel} \left(M^2+(t-u) m_A^2\right) m_b \phi
   _{A_\parallel}(v)\right)\\\nonumber&+\frac{3m_A}{2 M^4} \int
   _0^{s_0} ds\int _0^u du\int _0^1 dt\int _0^u dv\frac{e^{-\frac{s}{M^2}} \delta
   \left(s-s\left(t,q^2,m_A^2\right)\right)}{t^2 (v-1)}\\\nonumber&\qquad\qquad \times \left(t (1-2 u) f_{A_\perp} m_A \phi
   _{A_\perp}(v) M^2+f_{A_\parallel} \left(M^2+(t-u) m_A^2\right) m_b \phi
   _{A_\parallel}(v)\right) \\\nonumber&-\frac{3 m_A}{2
   M^4}\int
   _0^{s_0} ds\int _0^1 du\int _0^u dt\int _u^1 dv\frac{e^{-\frac{s}{M^2}} \delta
   \left(s-s\left(t,q^2,m_A^2\right)\right)}{t^2 v}\\\nonumber&\qquad\qquad \times \left(t (1-2 u) f_{A_\perp} m_A \phi
   _{A_\perp}(v) M^2+f_{A_\parallel} \left(M^2+(t-u) m_A^2\right) m_b \phi
   _{A_\parallel}(v)\right)\\\nonumber&+3f_{A_\parallel} m_b m_A^3\dfrac{\partial}{\partial s_0}  e^{-\frac{s_0}{M^2}}
   \int _0^1 du\int _0^u dt\frac{(t-u) \delta \left(s_0-s\left(t,q^2,m_A^2\right)\right)
   \phi _{A_\parallel}(u)}{t^2} \\\nonumber&+\frac{3f_{A_\parallel} m_b m_A^3}{2}\dfrac{\partial}{\partial s_0}
   e^{-\frac{s_0}{M^2}} \int _0^1 du\int _0^u dt\int _0^u dv\frac{(t-u) \delta
   \left(s_0-s\left(t,q^2,m_A^2\right)\right) \phi _{A_\parallel}(v)}{t^2
   (v-1)} \\\nonumber&+\frac{3f_{A_\parallel} m_b m_A^3}{2}\dfrac{\partial}{\partial s_0} e^{-\frac{s_0}{M^2}}
   \int _0^1 du\int _0^u dt\int _u^1 dv\frac{(u-t) \delta
   \left(s_0-s\left(t,q^2,m_A^2\right)\right) \phi _{A_\parallel}(v)}{t^2 v}
   \label{eq:eq46}
   \end{align}
   
\begin{align}
	f^n_{V_-}(q^2)\kappa_{f^n_{V_-}} e^{-\frac{m_{B_c}^2}{M^2}}&=-\frac{3f_{A_\perp}}{2} \int _0^{s_0} ds\int _0^1 du\frac{e^{-\frac{s}{M^2}} \delta
   \left(s-s\left(u,q^2,m_A^2\right)\right) \phi _{A_\perp}(u)}{u}
   \\\nonumber&-\frac{3 e^{-\frac{s_0}{M^2}} m_A}{2 M^2}\int _0^1 du\int _0^u dt\frac{\delta
   \left(s_0-s\left(t,q^2,m_A^2\right)\right)}{t^3} \\\nonumber&\qquad\qquad \times\left((t-2) t f_{A_\perp} m_A \phi
   _{A_\perp}(u) M^2+2 f_{A_\parallel} \left(t M^2+(t-2) (t-u) m_A^2\right) m_b \phi
   _{A_\parallel}(u)\right)\\\nonumber&-\frac{3m_A}{2
   M^4} \int _0^{s_0} ds\int
   _0^1 du\int _0^u dt\frac{e^{-\frac{s}{M^2}} \delta \left(s-s\left(t,q^2,m_A^2\right)\right)}{t^3}\\\nonumber&\qquad\qquad \times
   \left((t-2) t f_{A_\perp} m_A \phi _{A_\perp}(u) M^2+2 f_{A_\parallel} \left(t
   M^2+(t-2) (t-u) m_A^2\right) m_b \phi _{A_\parallel}(u)\right)\\\nonumber&-\frac{3 e^{-\frac{s_0}{M^2}}m_A}{2 M^2} \int _0^1 du\int _0^u dt\int _0^u dv\frac{\delta
   \left(s_0-s\left(t,q^2,m_A^2\right)\right)}{t^3 (v-1)}\\\nonumber&\qquad\qquad \times \bigg(t (-2 u t+t+4 u-2) f_{A_\perp} m_A
   \phi _{A_\perp}(v) M^2 \\\nonumber&\qquad\qquad\quad\quad+f_{A_\parallel} \left(t M^2+(t-2) (t-u) m_A^2\right) m_b \phi
   _{A_\parallel}(v)\bigg) \\\nonumber&+\frac{3
   e^{-\frac{s_0}{M^2}}m_A}{2 M^2} \int _0^1 du\int _0^u dt\int _u^1 dv\frac{\delta
   \left(s_0-s\left(t,q^2,m_A^2\right)\right)}{t^3 v} \\\nonumber&\qquad\qquad \times\bigg(t (-2 u t+t+4 u-2) f_{A_\perp} m_A
   \phi _{A_\perp}(v) M^2+\\\nonumber&\qquad\qquad\quad\quad f_{A_\parallel} \left(t M^2+(t-2) (t-u) m_A^2\right) m_b \phi
   _{A_\parallel}(v)\bigg) \\\nonumber&-\frac{3m_A}{2 M^4} \int
   _0^{s_0} ds\int _0^1 du\int _0^u dt\int _0^u dv\frac{e^{-\frac{s}{M^2}} \delta
   \left(s-s\left(t,q^2,m_A^2\right)\right)}{t^3 (v-1)}\\\nonumber&\qquad\qquad \times \bigg(t (-2 u t+t+4 u-2) f_{A_\perp} m_A
   \phi _{A_\perp}(v) M^2+\\\nonumber&\qquad\qquad\quad\quad f_{A_\parallel} \left(t M^2+(t-2) (t-u) m_A^2\right) m_b \phi
   _{A_\parallel}(v)\bigg) \\\nonumber&+\frac{3m_A}{2
   M^4} \int
   _0^{s_0} ds\int _0^1 du\int _0^u dt\int _u^1 dv\frac{e^{-\frac{s}{M^2}} \delta
   \left(s-s\left(t,q^2,m_A^2\right)\right)}{t^3 v}\\\nonumber&\qquad\qquad \times \bigg(t (-2 u t+t+4 u-2) f_{A_\perp} m_A
   \phi _{A_\perp}(v) M^2+\\\nonumber&\qquad\qquad\quad\quad f_{A_\parallel} \left(t M^2+(t-2) (t-u) m_A^2\right) m_b \phi
   _{A_\parallel}(v)\bigg)\\\nonumber&+ -3f_{A_\parallel} m_b m_A^3 \dfrac{\partial}{\partial s_0}e^{-\frac{s_0}{M^2}}
   \int _0^1 du\int _0^u dt\frac{(t-2) (t-u) \delta
   \left(s_0-s\left(t,q^2,m_A^2\right)\right) \phi _{A_\parallel}(u)}{t^3}
   \\\nonumber&-\frac{3f_{A_\parallel} m_b m_A^3}{2} \dfrac{\partial}{\partial s_0}e^{-\frac{s_0}{M^2}} \int _0^1 du\int _0^u dt\int
   _0^u dv\frac{(t-2) (t-u) \delta \left(s_0-s\left(t,q^2,m_A^2\right)\right) \phi
   _{A_\parallel}(v)}{t^3 (v-1)} \\\nonumber&+\frac{3f_{A_\parallel} m_b m_A^3}{2}\dfrac{\partial}{\partial s_0}
   e^{-\frac{s_0}{M^2}} \int _0^1 du\int _0^u dt\int _u^1 dv\frac{(t-2) (t-u) \delta
   \left(s_0-s\left(t,q^2,m_A^2\right)\right) \phi _{A_\parallel}(v)}{t^3 v} 
   \label{eq:eq60}
\end{align}  
  
   
    \begin{align}
   	f^n_{T}(q^2)\kappa_{f^n_{T}} e^{-\frac{m_{B_c}^2}{M^2}}&=\frac{3}{2}f^n_{T_\perp} \int _0^1du\,\int _0^u dt\,\frac{ e^{-\frac{s_0}{M^2}} 
   m_{T}  \delta \left(s_0-s(t;q^2,m_T^2)\right) \phi^n _{T_\perp}(u)}{t^2}\\ \nonumber&+\frac{3}{2}f^n_{T_\perp}\int
   _0^{s_0}ds\,e^{-\frac{s}{M^2}}\int _0^1du\,\int _0^u dt\,\frac{  
    m_{T}  \delta
   \left(s-s(t;q^2,m_T^2)\right) \phi^n
   _{T_\perp}(u)}{M^2 t^2 }
   \label{eq:eq48}
   \end{align}
   
   \begin{align}
   	f^n_{T_0}(q^2)\kappa_{f^n_{T_0}} e^{-\frac{m_{B_c}^2}{M^2}}&=\frac{3}{2}f^n_{T_\perp}\int _0^1 du\,\int _0^u dt\,\frac{ e^{-\frac{s_0}{M^2}} 
   m_{T} \left(m_b^2+t^2 m_{T}^2-q^2\right) \delta \left(s_0-s(t;q^2,m_T^2)\right) \phi^n _{T_\perp}(u)}{t^3 }\\ \nonumber&+\frac{3}{2}f^n_{T_\perp}\int _0^{s_0}ds\,e^{-\frac{s}{M^2}}\int _0^1du\,\int _0^udt\,\frac{
     m_{T}  \delta \left(s-s(t;q^2,m_T^2)\right)}{M^2 t^3 }\\ \nonumber &\qquad\qquad\times \left(-t
   M^2+m_b^2+t^2 m_{T}^2-q^2\right)\phi^n _{T_\perp}(u)
   \label{eq:eq49}
   \end{align}
   
   \begin{align}
   	f^n_{T_+}(q^2)\kappa_{f^n_{T_+}} e^{-\frac{m_{B_c}^2}{M^2}}&=-6f^n_{T_\parallel}\dfrac{\partial}{\partial s_0}\int _0^1 du\,\int _0^u dt\,\frac{ e^{-\frac{s_0}{M^2}}  m_{T}^2
   \delta \left(s_0-s(t;q^2,m_T^2)\right)  (t-u)
     m_b  \phi3n _{T_\parallel}(u)}{ t^3 }\\ \nonumber&-3\int _0^1 du\,\int _0^u dt\,\frac{ e^{-\frac{s_0}{M^2}}  m_{T}^2
   \delta \left(s_0-s(t;q^2,m_T^2)\right)}{M^2 t^3 }\\ \nonumber &\qquad\qquad\times \left(t
   f^n_{T_\perp} \phi^n _{T_\perp}(u) M^2+2 (t-u)
   f^n_{T_\parallel} m_b  \phi^n _{T_\parallel}(u)\right)\\ \nonumber &-6\dfrac{\partial}{\partial s_0}\int _0^1 du\,\int _0^u dt\,\int
   _0^u dv\,\frac{ e^{-\frac{s_0}{M^2}} (t-u)
    f^n_{T_\parallel} m_b
    m_{T}^2 \delta \left(s_0-s(t;q^2,m_T^2)\right) \phi^n _{T_\parallel}(v)}{ t^3 (v-1)
   }\\ \nonumber&-6\int _0^1 du\,\int _0^u dt\,\int
   _0^u dv\,\frac{ e^{-\frac{s_0}{M^2}} (t-u)
    f^n_{T_\parallel} m_b
    m_{T}^2 \delta \left(s_0-s(t;q^2,m_T^2)\right) \phi^n _{T_\parallel}(v)}{M^2 t^3 (v-1)
   }\\ \nonumber &+6\dfrac{\partial}{\partial s_0}\int _0^1 du\,\int _0^u dt\,\int
   _u^1 dv\,\frac{ e^{-\frac{s_0}{M^2}} (t-u)
    f^n_{T_\parallel} m_b
    m_{T}^2 \delta \left(s_0-s(t;q^2,m_T^2)\right) \phi^n _{T_\parallel}(v)}{ t^3 v
   }\\ \nonumber&+6\int _0^1 du\,\int _0^u dt\,\int
   _u^1 dv\,\frac{ e^{-\frac{s_0}{M^2}} (t-u)
    f^n_{T_\parallel} m_b
    m_{T}^2 \delta \left(s_0-s(t;q^2,m_T^2)\right) \phi^n _{T_\parallel}(v)}{M^2 t^3 v
   }\\ \nonumber &-3\int
   _0^{s_0} ds\,e^{-\frac{s}{M^2}}\int _0^1 du\,\int _0^u dt\, \frac{   m_{T} \delta \left(s-s(t;q^2,m_T^2)\right)}{M^4 t^3 }\\ \nonumber
   &\qquad\qquad\times\left(t f^n_{T_\perp} \phi^n _{T_\perp}(u) M^2+2 (t-u) f^n_{T_\parallel} m_b m_{T} \phi^n
   _{T_\parallel}(u)\right)\\ \nonumber &-6\int _0^{s_0} ds\,e^{-\frac{s}{M^2}} \int _0^1 du \,\int _0^u dt\,\int
   _0^u dv\, \frac{  (t-u) f^n_{T_\parallel} m_b  m_{T}^2 \delta \left(s-s(t;q^2,m_T^2)\right)
   \phi^n _{T_\parallel}(v)}{M^4 t^3 (v-1) }\\ \nonumber &+6\int _0^{s_0} ds\,\int _0^1 \,du\int _0^u dt\,\int
   _u^1 dv\,\frac{ e^{-\frac{s}{M^2}} (t-u) f^n_{T_\parallel}  m_{T}^2 \delta \left(s-s(t;q^2,m_T^2)\right)
   \phi^n _{T_\parallel}(v)}{M^4 t^3 v }
   \label{eq:eq50}
   \end{align}

\begin{equation}
     f^n_{T_-}=- f^n_{T_+}
\end{equation}

\section{Fit Parameters}
\label{app:fpexp}

\begin{table}[h] \centering 
\footnotesize
\begin{tabular}{||l|lcc||}
\hline
\multicolumn{4}{|c|}{$\boldsymbol{f^n_{1}}$}\\
\hline
\textbf{State}& $\boldsymbol{a}$ & $\boldsymbol{b}$ & $\boldsymbol{m_{fit}\,(GeV)}$\\ 
\hline
$n=1$ & $1.40$ & $-1.07$ & $5.18$ \\
\hline
$n=2$ & $0.749$ & $-0.528$ & $5.18$ \\
\hline
$n=3$ & $0.388$ & $-0.260$ & $5.18$ \\
\hline
\end{tabular}
\begin{tabular}{||l|lcc||}
\hline
\multicolumn{4}{|c|}{$\boldsymbol{f^n_{2}}$}\\
\hline
\textbf{State}& $\boldsymbol{a}$ & $\boldsymbol{b}$ & $\boldsymbol{m_{fit}\,(GeV)}$\\ 
\hline
$n=1$ & $-1.55$ & $1.18$ & $5.18$ \\
\hline
$n=2$ & $-0.757$ & $0.533$ & $5.18$ \\
\hline
$n=3$ & $-0.378$ & $0.253$ & $5.18$ \\
\hline
\end{tabular}
\caption{Fit parameters of $f^n_1$ and $f^n_2$ form factors  for $^3P_0$ states in \cref{eq:eqfit}.}
\label{table:tab1}	
\end{table}
\begin{table}[h]\centering
\footnotesize
\begin{tabular}{||l|lcc||}
\hline
\multicolumn{4}{|c|}{$\boldsymbol{f^n_{V}}$}\\
\hline
\textbf{State}& $\boldsymbol{a}$ & $\boldsymbol{b}$ & $\boldsymbol{m_{fit} \,(GeV)}$\\
\hline
$n=1$ & $-1.39$ & $1.14$ & $5.18$ \\
\hline
$n=2$ & $-1.22$ & $0.967$ & $5.18$ \\
\hline
$n=3$ & $-1.52$ & $1.12$ & $5.18$ \\
\hline
\end{tabular}
\begin{tabular}{||l|lcc||}
\hline
\multicolumn{4}{|c|}{$\boldsymbol{f^n_{V_+} }$}\\
\hline
\textbf{State}& $\boldsymbol{a}$ & $\boldsymbol{b}$ & $\boldsymbol{m_{fit} \,(GeV)}$\\
\hline
$n=1$ & $1.97$ & $-1.56$ & $5.18$ \\
\hline
$n=2$ & $2.33$ & $-1.80$ & $5.18$ \\
\hline
$n=3$ & $2.03$ & $-1.55$ & $5.18$ \\
\hline
\end{tabular}\\
\vspace{1mm}
\begin{tabular}{||l|lcc||}
\hline
\multicolumn{4}{|c|}{$\boldsymbol{f^n_{V_-}}$}\\
\hline
\textbf{State}& $\boldsymbol{a}$ & $\boldsymbol{b}$ & $\boldsymbol{m_{fit}\,(GeV)}$\\
\hline
$n=1$ & $-4.15$ & $3.28$ & $5.18$ \\
\hline
$n=2$ & $-3.39$ & $2.62$ & $5.18$ \\
\hline
$n=3$ & $-2.25$ & $1.71$ & $5.18$ \\
\hline
\end{tabular}
\begin{tabular}{||l|lcc||}
\hline
\multicolumn{4}{|c|}{$\boldsymbol{f^n_{V_0}}$}\\
\hline
\textbf{State}& $\boldsymbol{a}$ & $\boldsymbol{b}$ & $\boldsymbol{m_{fit}\,(GeV)}$\\
\hline
$n=1$ & $-0.301$ & $0.202$ & $5.18$ \\
\hline
$n=2$ & $-0.230$ & $0.149$ & $5.18$ \\
\hline
$n=3$ & $-0.241$ & $0.149$ & $5.18$ \\
\hline
\end{tabular}
\caption{Fit parameters of $f^n_{V}$, $f^n_{{V_+}}$, $f^n_{{V_-}}$, and $f^n_{{V_0}}$ form factors for $^1P_1$ states in \cref{eq:eqfit}.}
\label{table:tab2}
\end{table}
\begin{table}[h]\centering
\footnotesize
\begin{tabular}{||l|lcc||}
\hline
\multicolumn{4}{|c|}{$\boldsymbol{f^n_{V}}$}\\
\hline
\textbf{State}& $\boldsymbol{a}$ & $\boldsymbol{b}$ & $\boldsymbol{m_{fit}\,(GeV)}$\\
\hline
$n=1$ & $4.85$ & $-3.80$ & $5.18$ \\
\hline
$n=2$ & $3.67$ & $-2.70$ & $5.18$ \\
\hline
$n=3$ & $2.15$ & $-1.50$ & $5.18$ \\
\hline
\end{tabular}
\begin{tabular}{||l|lcc||}
\hline
\multicolumn{4}{|c|}{$\boldsymbol{f^n_{V_+}}$}\\
\hline
\textbf{State}& $\boldsymbol{a}$ & $\boldsymbol{b}$ & $\boldsymbol{m_{fit}\,(GeV)}$\\
\hline
$n=1$ & $-1.22$ & $0.897$ & $5.18$ \\
\hline
$n=2$ & $-0.457$ & $0.372$ & $5.18$ \\
\hline
$n=3$ & $-0.0843$ & $0.110$ & $5.18$ \\
\hline
\end{tabular}\\
\vspace{1mm}
\begin{tabular}{||l|lcc||}
\hline
\multicolumn{4}{|c|}{$\boldsymbol{f^n_{V_-}}$}\\
\hline
\textbf{State}& $\boldsymbol{a}$ & $\boldsymbol{b}$ & $\boldsymbol{m_{fit}\,(GeV)}$\\
\hline
$n=1$ & $0.852$ & $-0.627$ & $5.18$ \\
\hline
$n=2$ & $0.492$ & $-0.400$ & $5.18$ \\
\hline
$n=3$ & $0.0895$ & $-0.116$ & $5.18$ \\
\hline
\end{tabular}
\begin{tabular}{||l|lcc||}
\hline
\multicolumn{4}{|c|}{$\boldsymbol{f^n_{V_0}}$}\\
\hline
\textbf{State}& $\boldsymbol{a}$ & $\boldsymbol{b}$ & $\boldsymbol{m_{fit}\,(GeV)}$\\
\hline
$n=1$ & $0.211$ & $-0.146$ & $5.18$ \\
\hline
$n=2$ & $0.0933$ & $-0.0598$ & $5.18$ \\
\hline
$n=3$ & $0.149$ & $-0.0907$ & $5.18$ \\
\hline
\end{tabular}
\caption{Fit parameters of $f^n_{V}$, $f^n_{{V_+}}$, $f^n_{{V_-}}$, and $f^n_{{V_0}}$ form factors for $^3P_1$ states in \cref{eq:eqfit}.}	
\label{table:tab3}	
\end{table}
\begin{table}\centering
\footnotesize
\begin{tabular}{||l|lcc||}
\hline
\multicolumn{4}{|c|}{$\boldsymbol{f^n_{T}}$}\\
\hline
\textbf{State}& $\boldsymbol{a}$ & $\boldsymbol{b}$ & $\boldsymbol{m_{fit}\,(GeV)}$\\
\hline
$n=1$ & $-0.765$ & $0.646$ & $5.18$ \\
\hline
$n=2$ & $-0.856$ & $0.698$ & $5.18$ \\
\hline
$n=3$ & $-0.721$ & $0.573$ & $5.18$ \\
\hline
\end{tabular}
\begin{tabular}{||l|lcc||}
\hline
\multicolumn{4}{|c|}{$\boldsymbol{f^n_{T_+}=-f^n_{T_-}}$}\\
\hline
\textbf{State}& $\boldsymbol{a}$ & $\boldsymbol{b}$ & $\boldsymbol{m_{fit}\,(GeV)}$\\
\hline
$n=1$ & $0.400$ & $-0.320$ & $5.18$ \\
\hline
$n=2$ & $0.428$ & $-0.334$ & $5.18$ \\
\hline
$n=3$ & $0.360$ & $-0.274$ & $5.18$ \\
\hline
\end{tabular}\\
\vspace{1mm}
\begin{tabular}{||l|lcc||}
\hline
\multicolumn{4}{|c|}{$\boldsymbol{f^n_{T_0}}$}\\
\hline
\textbf{State}& $\boldsymbol{a}$ & $\boldsymbol{b}$ & $\boldsymbol{m_{fit}\,(GeV)}$\\
\hline
$n=1$ & $-0.692$ & $0.526$ & $5.18$ \\
\hline
$n=2$ & $-0.790$ & $-0.588$ & $5.18$ \\
\hline
$n=3$ & $-0.300$ & $0.218$ & $5.18$ \\
\hline
\end{tabular}
\caption{Fit parameters of $f^n_{T}$, $f^n_{T_+}$, $f^n_{T_-}$, and $f^n_{T_0}$ form factors for $^3P_2$ states in \cref{eq:eqfit}.}		
\label{table:tab4}
\end{table}

\begin{table}[h] \centering 
\footnotesize
\begin{tabular}{||l|lccc||}
\hline
\multicolumn{5}{|c|}{$\boldsymbol{f^n_{1}}$}\\
\hline
\textbf{State}& $\boldsymbol{F_0}$ &$\boldsymbol{a}$ & $\boldsymbol{b}$ & $\boldsymbol{m_{fit}\,(GeV)}$\\ 
\hline
$n=1$ & $0.323$ &$2.98$ & $-0.514$ & $7.51$ \\
\hline
$n=2$ & $0.210$ &$1.40$ & $-2.03$ & $7.05$ \\
\hline
$n=3$ & $0.119$ &$0.407$ & $0.0532$ & $5.12$ \\
\hline
\end{tabular}
\begin{tabular}{||l|lccc||}
\hline
\multicolumn{5}{|c|}{$\boldsymbol{f^n_{2}}$}\\
\hline
\textbf{State}& $\boldsymbol{F_0}$ &$\boldsymbol{a}$ & $\boldsymbol{b}$ & $\boldsymbol{m_{fit}\,(GeV)}$\\ 
\hline
$n=1$ & $-0.332$ &$2.49$ & $0.262$ & $5.84$ \\
\hline
$n=2$ & $-0.228$ &$3.22$ & $0.682$ & $8.01$ \\
\hline
$n=3$ & $-0.139$ &$1.43$ & $0.110$ & $6.68$ \\
\hline
\end{tabular}
\caption{Fit parameters of $f^n_1$ and $f^n_2$ form factors  for $^3P_0$ states in \cref{eq:expfit}.}
\label{table:tab6}	
\end{table}
\begin{table}[h]\centering
\footnotesize
\begin{tabular}{||l|lccc||}
\hline
\multicolumn{5}{|c|}{$\boldsymbol{f^n_{V}}$}\\
\hline
\textbf{State}& $\boldsymbol{F_0}$ &$\boldsymbol{a}$ & $\boldsymbol{b}$ & $\boldsymbol{m_{fit} \,(GeV)}$\\
\hline
$n=1$ & $0.229$ &$6.91$ & $7.06$ & $8.58$ \\
\hline
$n=2$ & $0.231$ &$2.02$ & $0.348$ & $5.10$ \\
\hline
$n=3$ & $0.368$ &$3.69$ & $2.75$ & $8.68$ \\
\hline
\end{tabular}
\begin{tabular}{||l|lccc||}
\hline
\multicolumn{5}{|c|}{$\boldsymbol{f^n_{V_+} }$}\\
\hline
\textbf{State}& $\boldsymbol{F_0}$ &$\boldsymbol{a}$ & $\boldsymbol{b}$ & $\boldsymbol{m_{fit} \,(GeV)}$\\
\hline
$n=1$ & $0.291$ &$2.41$ & $-1.82$ & $9.53$ \\
\hline
$n=2$ & $0.274$ &$1.44$ & $-2.08$ & $8.71$ \\
\hline
$n=3$ & $0.431$ &$1.32$ & $5.93$ & $6.61$ \\
\hline
\end{tabular}\\
\vspace{1mm}
\begin{tabular}{||l|lccc||}
\hline
\multicolumn{5}{|c|}{$\boldsymbol{f^n_{V_-}}$}\\
\hline
\textbf{State}& $\boldsymbol{F_0}$ &$\boldsymbol{a}$ & $\boldsymbol{b}$ & $\boldsymbol{m_{fit} \,(GeV)}$\\
\hline
$n=1$ & $0.397$ &$4.07$ & $3.89$ & $6.28$ \\
\hline
$n=2$ & $0.418$ &$2.45$ & $1.97$ & $4.85$ \\
\hline
$n=3$ & $0.530$ &$2.89$ & $1.60$ & $6.11$ \\
\hline
\end{tabular}
\begin{tabular}{||l|lccc||}
\hline
\multicolumn{5}{|c|}{$\boldsymbol{f^n_{V_0}}$}\\
\hline
\textbf{State}& $\boldsymbol{F_0}$ &$\boldsymbol{a}$ & $\boldsymbol{b}$ & $\boldsymbol{m_{fit} \,(GeV)}$\\
\hline
$n=1$ & $-0.141$ &$2.70$ & $4.69$ & $8.94$ \\
\hline
$n=2$ & $-0.106$ &$2.23$ & $4.05$ & $8.54$ \\
\hline
$n=3$ & $-0.116$ &$2.23$ & $2.24$ & $7.71$ \\
\hline
\end{tabular}
\caption{Fit parameters of $f^n_{V}$, $f^n_{{V_+}}$, $f^n_{{V_-}}$, and $f^n_{{V_0}}$ form factors for $^1P_1$ states in \cref{eq:expfit}.}
\label{table:tab7}
\end{table}
\begin{table}[h]\centering
\footnotesize
\begin{tabular}{||l|lccc||}
\hline
\multicolumn{5}{|c|}{$\boldsymbol{f^n_{V}}$}\\
\hline
\textbf{State}& $\boldsymbol{F_0}$ &$\boldsymbol{a}$ & $\boldsymbol{b}$ & $\boldsymbol{m_{fit} \,(GeV)}$\\
\hline
$n=1$ & $-0.966$ &$4.63$ & $3.03$ & $8.01$ \\
\hline
$n=2$ & $-0.918$ &$1.81$ & $-0.625$ & $6.20$ \\
\hline
$n=3$ & $-0.621$ &$1.68$ & $0.00820$ & $7.51$ \\
\hline
\end{tabular}
\begin{tabular}{||l|lccc||}
\hline
\multicolumn{5}{|c|}{$\boldsymbol{f^n_{V_+} }$}\\
\hline
\textbf{State}& $\boldsymbol{F_0}$ &$\boldsymbol{a}$ & $\boldsymbol{b}$ & $\boldsymbol{m_{fit} \,(GeV)}$\\
\hline
$n=1$ & $-0.666$ &$2.12$ & $0.955$ & $6.00$ \\
\hline
$n=2$ & $-0.633$ &$2.34$ & $0.157$ & $7.08$ \\
\hline
$n=3$ & $-0.476$ &$1.95$ & $0.741$ & $7.45$ \\
\hline
\end{tabular}\\
\vspace{1mm}
\begin{tabular}{||l|lccc||}
\hline
\multicolumn{5}{|c|}{$\boldsymbol{f^n_{V_-}}$}\\
\hline
\textbf{State}& $\boldsymbol{F_0}$ &$\boldsymbol{a}$ & $\boldsymbol{b}$ & $\boldsymbol{m_{fit} \,(GeV)}$\\
\hline
$n=1$ & $0.249$ &$2.02$ & $0.213$ & $5.64$ \\
\hline
$n=2$ & $0.180$ &$0.976$ & $-6.27$ & $6.80$ \\
\hline
$n=3$ & $0.0320$ &$-3.37$ & $-5.93$ & $4.95$ \\
\hline
\end{tabular}
\begin{tabular}{||l|lccc||}
\hline
\multicolumn{5}{|c|}{$\boldsymbol{f^n_{V_0}}$}\\
\hline
\textbf{State}& $\boldsymbol{F_0}$ &$\boldsymbol{a}$ & $\boldsymbol{b}$ & $\boldsymbol{m_{fit} \,(GeV)}$\\
\hline
$n=1$ & $0.0501$ &$2.86$ & $6.18$ & $7.58$ \\
\hline
$n=2$ & $0.0234$ &$2.20$ & $1.40$ & $5.00$ \\
\hline
$n=3$ & $0.0343$ &$2.73$ & $-0.808$ & $5.05$ \\
\hline
\end{tabular}
\caption{Fit parameters of $f^n_{V}$, $f^n_{{V_+}}$, $f^n_{{V_-}}$, and $f^n_{{V_0}}$ form factors for $^3P_1$ states in \cref{eq:expfit}.}
\label{table:tab8}
\end{table}
\begin{table}[h]\centering
\footnotesize
\begin{tabular}{||l|lccc||}
\hline
\multicolumn{5}{|c|}{$\boldsymbol{f^n_{T}}$}\\
\hline
\textbf{State}& $\boldsymbol{F_0}$ &$\boldsymbol{a}$ & $\boldsymbol{b}$ & $\boldsymbol{m_{fit} \,(GeV)}$\\
\hline
$n=1$ & $-0.118$ &$5.57$ & $3.36$ & $6.93$ \\
\hline
$n=2$ & $-0.157$ &$4.22$ & $1.87$ & $6.62$ \\
\hline
$n=3$ & $-0.151$ &$2.90$ & $0.224$ & $6.17$ \\
\hline
\end{tabular}
\begin{tabular}{||l|lccc||}
\hline
\multicolumn{5}{|c|}{$\boldsymbol{f^n_{T_+}=-f^n_{T_-} }$}\\
\hline
\textbf{State}& $\boldsymbol{F_0}$ &$\boldsymbol{a}$ & $\boldsymbol{b}$ & $\boldsymbol{m_{fit} \,(GeV)}$\\
\hline
$n=1$ & $0.0818$ &$2.85$ & $1.33$ & $5.82$ \\
\hline
$n=2$ & $0.0985$ &$2.88$ & $0.343$ & $6.38$ \\
\hline
$n=3$ & $-0.0640$ &$1.94$ & $1.55$ & $7.08$ \\
\hline
\end{tabular}\\
\vspace{1mm}
\begin{tabular}{||l|lccc||}
\hline
\multicolumn{5}{|c|}{$\boldsymbol{f^n_{T_0}}$}\\
\hline
\textbf{State}& $\boldsymbol{F_0}$ &$\boldsymbol{a}$ & $\boldsymbol{b}$ & $\boldsymbol{m_{fit} \,(GeV)}$\\
\hline
$n=1$ & $-0.170$ &$2.86$ & $2.29$ & $6.96$ \\
\hline
$n=2$ & $-0.209$ &$2.75$ & $1.55$ & $7.30$ \\
\hline
$n=3$ & $-0.102$ &$2.39$ & $1.52$ & $7.38$ \\
\hline
\end{tabular}
\caption{Fit parameters of $f^n_{T}$, $f^n_{{T_+}}$, $f^n_{{T_-}}$, and $f^n_{{T_0}}$ form factors for $^3P_2$ states in \cref{eq:expfit}.}
\label{table:tab9}
\end{table}

\end{appendix}


\newpage
\begin{thebibliography}{200}
\bibliographystyle{ieeetr}

\bibitem{LHCb:2014rck}
R.~Aaij \textit{et al.} [LHCb],
Phys. Rev. D \textbf{90} (2014) no.3, 032009
doi:10.1103/PhysRevD.90.032009
[arXiv:1407.2126 [hep-ex]].

\bibitem{LHCb:2017vlu}
R.~Aaij \textit{et al.} [LHCb],
Phys. Rev. Lett. \textbf{120} (2018) no.12, 121801
doi:10.1103/PhysRevLett.120.121801
[arXiv:1711.05623 [hep-ex]].

\bibitem{Godfrey:1985xj}
S.~Godfrey and N.~Isgur,
Phys. Rev. D \textbf{32} (1985), 189-231
doi:10.1103/PhysRevD.32.189

\bibitem{CDF:1998axz}
F.~Abe \textit{et al.} [CDF],
Phys. Rev. D \textbf{58} (1998), 112004
doi:10.1103/PhysRevD.58.112004
[arXiv:hep-ex/9804014 [hep-ex]].
 
\bibitem{Aliev:1999tg}
T.~Aliev and M.~Savci,
Phys. Lett. B \textbf{480}, 97-104 (2000)
doi:10.1016/S0370-2693(00)00378-6
[arXiv:hep-ph/9908203 [hep-ph]].

\bibitem{Aliev:2006vs}
T.~Aliev and M.~Savci,
Eur. Phys. J. C \textbf{47}, 413-421 (2006)
doi:10.1140/epjc/s2006-02579-5
[arXiv:hep-ph/0601267 [hep-ph]].

\bibitem{Azizi:2007jx}
K.~Azizi and V.~Bashiry,
Phys. Rev. D \textbf{76}, 114007 (2007)
doi:10.1103/PhysRevD.76.114007
[arXiv:0708.2068 [hep-ph]].

\bibitem{Kang:2018jzg}
X.~W.~Kang, T.~Luo, Y.~Zhang, L.~Y.~Dai and C.~Wang,
Eur. Phys. J. C \textbf{78} (2018) no.11, 909
doi:10.1140/epjc/s10052-018-6385-9
[arXiv:1808.02432 [hep-ph]].

\bibitem{Ghahramany:2008tz}
N.~Ghahramany, R.~Khosravi and K.~Azizi,
Phys. Rev. D \textbf{78}, 116009 (2008)
doi:10.1103/PhysRevD.78.116009
[arXiv:0811.2674 [hep-ph]].

\bibitem{Azizi:2008vv}
K.~Azizi, F.~Falahati, V.~Bashiry and S.~Zebarjad,
Phys. Rev. D \textbf{77}, 114024 (2008)
doi:10.1103/PhysRevD.77.114024
[arXiv:0806.0583 [hep-ph]].

\bibitem{Aliev:1998mq}
T.~Aliev and M.~Savci,
Phys. Lett. B \textbf{434}, 358-364 (1998)
doi:10.1016/S0370-2693(98)00692-3
[arXiv:hep-ph/9804407 [hep-ph]].

\bibitem{Aliev:1998ka}
T.~Aliev and M.~Savci,
J. Phys. G \textbf{24}, 2223-2228 (1998)
doi:10.1088/0954-3899/24/12/006
[arXiv:hep-ph/9805239 [hep-ph]].
 
\bibitem{Azizi:2008tw}
K.~Azizi, R.~Khosravi and V.~Bashiry,
Eur. Phys. J. C \textbf{56}, 357-370 (2008)
doi:10.1140/epjc/s10052-008-0668-5
[arXiv:0805.2806 [hep-ph]].

\bibitem{Azizi:2008vy}
K.~Azizi and R.~Khosravi,
Phys. Rev. D \textbf{78}, 036005 (2008)
doi:10.1103/PhysRevD.78.036005
[arXiv:0806.0590 [hep-ph]].

\bibitem{Du:1988ws}
D.~s.~Du and Z.~Wang,
Phys. Rev. D \textbf{39} (1989), 1342
doi:10.1103/PhysRevD.39.1342

\bibitem{Gershtein:1994jw}
S.~S.~Gershtein, V.~V.~Kiselev, A.~K.~Likhoded and A.~V.~Tkabladze,
Phys. Usp. \textbf{38} (1995), 1-37
doi:10.1070/PU1995v038n01ABEH000063
[arXiv:hep-ph/9504319 [hep-ph]].

\bibitem{Qiao:2012hp}
C.~F.~Qiao, P.~Sun, D.~Yang and R.~L.~Zhu,
Phys. Rev. D \textbf{89} (2014) no.3, 034008
doi:10.1103/PhysRevD.89.034008
[arXiv:1209.5859 [hep-ph]].

\bibitem{Esposito:2014rxa}
A.~Esposito, A.~L.~Guerrieri, F.~Piccinini, A.~Pilloni and A.~D.~Polosa,
Int. J. Mod. Phys. A \textbf{30} (2015), 1530002
doi:10.1142/S0217751X15300021
[arXiv:1411.5997 [hep-ph]].

\bibitem{Canham:2009zq}
D.~L.~Canham, H.~W.~Hammer and R.~P.~Springer,
Phys. Rev. D \textbf{80} (2009), 014009
doi:10.1103/PhysRevD.80.014009
[arXiv:0906.1263 [hep-ph]].

\bibitem{Sun:2015uva}
Z.~F.~Sun, M.~Bayar, P.~Fernandez-Soler and E.~Oset,
Phys. Rev. D \textbf{93} (2016) no.5, 054028
doi:10.1103/PhysRevD.93.054028
[arXiv:1510.06316 [hep-ph]].

\bibitem{Kang:2016jxw}
X.~W.~Kang and J.~A.~Oller,
Eur. Phys. J. C \textbf{77} (2017) no.6, 399
doi:10.1140/epjc/s10052-017-4961-z
[arXiv:1612.08420 [hep-ph]].



\bibitem{Shifman:1978bx}
M.~A.~Shifman, A.~I.~Vainshtein and V.~I.~Zakharov,
Nucl. Phys. B \textbf{147} (1979), 385-447
doi:10.1016/0550-3213(79)90022-1

\bibitem{Alkofer:2000wg}
R.~Alkofer and L.~von Smekal,
Phys. Rept. \textbf{353} (2001), 281
doi:10.1016/S0370-1573(01)00010-2
[arXiv:hep-ph/0007355 [hep-ph]].

\bibitem{Marciano:1977su}
W.~J.~Marciano and H.~Pagels,
Phys. Rept. \textbf{36} (1978), 137
doi:10.1016/0370-1573(78)90208-9

\bibitem{Bigi:1992su}
I.~I.~Y.~Bigi, N.~G.~Uraltsev and A.~I.~Vainshtein,
Phys. Lett. B \textbf{293} (1992), 430-436
[erratum: Phys. Lett. B \textbf{297} (1992), 477-477]
doi:10.1016/0370-2693(92)90908-M
[arXiv:hep-ph/9207214 [hep-ph]].

\bibitem{Kiselev:1999sc}
V.~V.~Kiselev, A.~K.~Likhoded and A.~I.~Onishchenko,
Nucl. Phys. B \textbf{569} (2000), 473-504
doi:10.1016/S0550-3213(99)00505-2
[arXiv:hep-ph/9905359 [hep-ph]].

\bibitem{Azizi:2009ny}
K.~Azizi, H.~Sundu and M.~Bayar,
Phys.\ Rev.\ D {\bf 79} (2009) 116001
doi:10.1103/PhysRevD.79.116001
[arXiv:0902.1467 [hep-ph]].


\bibitem{Cincioglu:2016fkm}
E.~Cincioglu, J.~Nieves, A.~Ozpineci and A.~U.~Yilmazer,
Eur. Phys. J. C \textbf{76} (2016) no.10, 576
doi:10.1140/epjc/s10052-016-4413-1
[arXiv:1606.03239 [hep-ph]].

\bibitem{Brodsky:1997de}
S.~J.~Brodsky, H.~C.~Pauli and S.~S.~Pinsky,
Phys. Rept. \textbf{301} (1998), 299-486
doi:10.1016/S0370-1573(97)00089-6
[arXiv:hep-ph/9705477 [hep-ph]].

\bibitem{Hwang:2009cu} 
  C.~W.~Hwang,
  JHEP {\bf 0910}, 074 (2009)
  doi:10.1088/1126-6708/2009/10/074
  [arXiv:0906.4412 [hep-ph]].
    
\bibitem{Olpak:2016wkf} 
M.~A.~Olpak, A.~Ozpineci and V.~Tanriverdi,
Phys.\ Rev.\ D {\bf 96}, no. 1, 014026 (2017)
doi:10.1103/PhysRevD.96.014026
[arXiv:1608.04539 [hep-ph]].

\bibitem{Leljak:2019eyw} 
D.~Leljak, B.~Melic and M.~Patra,
JHEP {\bf 1905}, 094 (2019)
doi:10.1007/JHEP05(2019)094
[arXiv:1901.08368 [hep-ph]].

\bibitem{Ebert:2010zu} 
D.~Ebert, R.~N.~Faustov and V.~O.~Galkin,
Phys.\ Rev.\ D {\bf 82}, 034019 (2010)
doi:10.1103/PhysRevD.82.034019
[arXiv:1007.1369 [hep-ph]].

\bibitem{Ivanov:2005fd}
M.~A.~Ivanov, J.~G.~Korner and P.~Santorelli,
Phys. Rev. D \textbf{71} (2005), 094006
[erratum: Phys. Rev. D \textbf{75} (2007), 019901]
doi:10.1103/PhysRevD.75.019901
[arXiv:hep-ph/0501051 [hep-ph]].

\bibitem{Faustov:2019mqr}
R.~N.~Faustov, V.~O.~Galkin and X.~W.~Kang,
Phys. Rev. D \textbf{101} (2020) no.1, 013004
doi:10.1103/PhysRevD.101.013004
[arXiv:1911.08209 [hep-ph]].

\bibitem{Issadykov:2017wlb}
A.~Issadykov, M.~A.~Ivanov and G.~Nurbakova,
EPJ Web Conf.\  {\bf 158} (2017) 03002
doi:10.1051/epjconf/201715803002
[arXiv:1907.13210 [hep-ph]].

\bibitem{Navarra:2000ji}
F.~S.~Navarra, M.~Nielsen, M.~E.~Bracco, M.~Chiapparini and C.~L.~Schat,
Phys. Lett. B \textbf{489} (2000), 319-328
doi:10.1016/S0370-2693(00)00967-9
[arXiv:hep-ph/0005026 [hep-ph]].

\bibitem{Bracco:2001dj}
M.~E.~Bracco, M.~Chiapparini, A.~Lozea, F.~S.~Navarra and M.~Nielsen,
Phys. Lett. B \textbf{521} (2001), 1-6
doi:10.1016/S0370-2693(01)01192-3
[arXiv:hep-ph/0108223 [hep-ph]].

\bibitem{Matheus:2002nq}
R.~D.~Matheus, F.~S.~Navarra, M.~Nielsen and R.~Rodrigues da Silva,
Phys. Lett. B \textbf{541} (2002), 265-272
doi:10.1016/S0370-2693(02)02259-1
[arXiv:hep-ph/0206198 [hep-ph]].

\bibitem{Navarra:2001ju}
F.~S.~Navarra, M.~Nielsen and M.~E.~Bracco,
Phys. Rev. D \textbf{65} (2002), 037502
doi:10.1103/PhysRevD.65.037502
[arXiv:hep-ph/0109188 [hep-ph]].

\bibitem{Bracco:2004rx}
M.~E.~Bracco, M.~Chiapparini, F.~S.~Navarra and M.~Nielsen,
Phys. Lett. B \textbf{605} (2005), 326-334
doi:10.1016/j.physletb.2004.11.024
[arXiv:hep-ph/0410071 [hep-ph]].

\bibitem{Carvalho:2005et}
F.~Carvalho, F.~O.~Duraes, F.~S.~Navarra and M.~Nielsen,
Phys. Rev. C \textbf{72} (2005), 024902
doi:10.1103/PhysRevC.72.024902
[arXiv:hep-ph/0508137 [hep-ph]].

\bibitem{Matheus:2005yu}
R.~D.~Matheus, F.~S.~Navarra, M.~Nielsen and R.~Rodrigues da Silva,
Int. J. Mod. Phys. E \textbf{14} (2005), 555-567
doi:10.1142/S0218301305003399

\bibitem{Rodrigues:2010ed}
B.~Osorio Rodrigues, M.~E.~Bracco, M.~Nielsen and F.~S.~Navarra,
Nucl. Phys. A \textbf{852} (2011), 127-140
doi:10.1016/j.nuclphysa.2011.01.001
[arXiv:1003.2604 [hep-ph]].

\bibitem{Bracco:2011pg}
M.~E.~Bracco, M.~Chiapparini, F.~S.~Navarra and M.~Nielsen,
Prog. Part. Nucl. Phys. \textbf{67} (2012), 1019-1052
doi:10.1016/j.ppnp.2012.03.002
[arXiv:1104.2864 [hep-ph]].

\bibitem{Bracco:2007sg}
M.~E.~Bracco, M.~Chiapparini, F.~S.~Navarra and M.~Nielsen,
Phys. Lett. B \textbf{659} (2008), 559-564
doi:10.1016/j.physletb.2007.11.066
[arXiv:0710.1878 [hep-ph]].

\bibitem{Zyla:2020zbs}
P.~A.~Zyla \textit{et al.} [Particle Data Group],
PTEP \textbf{2020} (2020) no.8, 083C01
doi:10.1093/ptep/ptaa104

\bibitem{Narison:2019tym}
S.~Narison,
Phys. Lett. B \textbf{802} (2020), 135221
doi:10.1016/j.physletb.2020.135221
[arXiv:1906.03614 [hep-ph]].


\bibitem{Hernandez:2006gt}
E.~Hernandez, J.~Nieves and J.~M.~Verde-Velasco,
Phys. Rev. D \textbf{74}, 074008 (2006)
doi:10.1103/PhysRevD.74.074008
[arXiv:hep-ph/0607150 [hep-ph]].


\bibitem{Ball:1996tb}
P.~Ball and V.~M.~Braun,
Phys. Rev. D \textbf{54} (1996), 2182-2193
doi:10.1103/PhysRevD.54.2182
[arXiv:hep-ph/9602323 [hep-ph]].

\bibitem{Braun:2000cs}
V.~M.~Braun and N.~Kivel,
Phys. Lett. B \textbf{501} (2001), 48-53
doi:10.1016/S0370-2693(01)00095-8
[arXiv:hep-ph/0012220 [hep-ph]].

\bibitem{Cheng:2010hn}
H.~Y.~Cheng, Y.~Koike and K.~C.~Yang,
Phys. Rev. D \textbf{82} (2010), 054019
doi:10.1103/PhysRevD.82.054019
[arXiv:1007.3541 [hep-ph]].

\bibitem{Wang:2009mi}
X.~X.~Wang, W.~Wang and C.~D.~Lu,
Phys. Rev. D \textbf{79} (2009), 114018
doi:10.1103/PhysRevD.79.114018
[arXiv:0901.1934 [hep-ph]].

\bibitem{Ivanov:2006ni}
M.~A.~Ivanov, J.~G.~Korner and P.~Santorelli,
Phys. Rev. D \textbf{73} (2006), 054024
doi:10.1103/PhysRevD.73.054024
[arXiv:hep-ph/0602050 [hep-ph]].

\bibitem{Ebert:2002pp}
D.~Ebert, R.~N.~Faustov and V.~O.~Galkin,
Phys. Rev. D \textbf{67} (2003), 014027
doi:10.1103/PhysRevD.67.014027
[arXiv:hep-ph/0210381 [hep-ph]].

  
\bibitem{Braguta:2008qe} 
  V.~V.~Braguta, A.~K.~Likhoded and A.~V.~Luchinsky,
  Phys.\ Rev.\ D {\bf 79}, 074004 (2009)
  doi:10.1103/PhysRevD.79.074004
  [arXiv:0810.3607 [hep-ph]].
  

\end{thebibliography}
\end{document}